\documentclass{article} % For LaTeX2e
\usepackage{times}
\usepackage{iclr2027_conference}

\usepackage{amsmath,amsfonts,bm}

\def\eqref#1{equation~\ref{#1}}
\def\1{\bm{1}}

\DeclareMathAlphabet{\mathsfit}{\encodingdefault}{\sfdefault}{m}{sl}
\SetMathAlphabet{\mathsfit}{bold}{\encodingdefault}{\sfdefault}{bx}{n}

\usepackage{hyperref}
\usepackage{url}
\usepackage{times}
\usepackage{latexsym}
\usepackage{booktabs}
\usepackage{multirow}
\usepackage{amsthm} 
\usepackage{amsmath,amsfonts,bm}
\usepackage{amssymb}        % for \varnothing
\usepackage{cleveref}
\usepackage{tabularx}
\usepackage{subcaption}
\usepackage{array}
\usepackage{algorithm}
\usepackage{algpseudocode}
\usepackage[table]{xcolor}
\usepackage{tcolorbox}
\tcbuselibrary{breakable,skins}
\usepackage{xcolor}
\definecolor{promptbg}{HTML}{F7F7F2}
\definecolor{promptrule}{HTML}{5B6770}
  \newcommand{\PH}[1]{\textit{\textcolor{purple}{$\langle$#1$\rangle$}}}
  
\newtcolorbox{promptbox}[1]{
  title=#1,
  colback=gray!5,
  fonttitle=\bfseries,
  breakable,
  enhanced,
  width=\linewidth,
  boxsep=1mm,
  left=1mm,
  right=1mm,
  top=1mm,
  bottom=1mm,
  before skip=6pt,
  after skip=6pt
}
  
\usepackage[T1]{fontenc}
\usepackage[utf8]{inputenc}

\usepackage{microtype}

\usepackage{inconsolata}

\usepackage{graphicx}

\newtheorem{definition}{Definition}

\title{Agents that Matter: How Subtle Choices Shape Agent Attribution in Multi-Agent Systems}

\author{Mingyu Lu, Yushan Huang, Chris Lin, Su-In Lee \\
Paul G. Allen School of Computer Science \& Engineering, University of Washington \\
\texttt{\{mingyulu, yushan13, clin25, suinlee\}@cs.washington.edu}
}

\iclrfinalcopy % Uncomment for camera-ready version, but NOT for submission.
\begin{document}

\maketitle

\begin{abstract}
As multi-agent systems (MAS) become increasingly complex, identifying the contributions of agents is critical for system optimization. However, existing approaches lack a unified formulation for credit assignment. In this work, we formalize agent attribution as a cooperative game, parameterized by the coalition distribution, intervention protocol, and target metric. Using this framework, we demonstrate that intervention protocols induce distinct games: Agent ablation isolates structural bottlenecks, whereas introspective LLM judges fail to faithfully approximate this behavior. We further find that Leave-One-Out (LOO) identifies bottleneck agents as effectively as combinatorial methods at much lower computational cost. Treating prompt optimization as an intervention protocol, we show that prompt-conditioned attribution can identify optimization targets and improve MAS performance. Finally, we apply our framework to audit a medical MAS, revealing that agent contributions to diagnostic accuracy and ethical behavior are often decoupled. By intervening on counterproductive roles, we observe an increase in ethics alignment while maintaining diagnostic accuracy. Overall, this work provides a principled approach for cost-effective MAS attribution and intervention. 

% Furthermore, to evaluate the utility of specific agent backbones, we introduce attribution via model replacement. By substituting underlying models of low-contribution agents, we improve task performance by up to 17\% while reducing cost by up to 35\% across three benchmarks. 

\end{abstract}

\vspace{-0.5em}
\section{Introduction}
\vspace{-0.5em}
Large language models (LLMs) are increasingly deployed as interactive agents \citep{wu2024autogen}, enabling collaborative multi-agent systems (MAS) that tackle complex tasks such as software engineering \citep{hong2023metagpt,qian2024chatdev} and medical diagnosis \citep{chen2025mdteamgpt}. Yet as these systems become more capable and complex, their performance becomes harder to trace back to specific agents: An agent's contribution depends not only on its underlying model capability but also on its assigned role and interactions with other agents \citep{zhu2025multiagentbench}. This creates a challenge for MAS evaluation and optimization: \emph{How can we quantify each agent’s contribution to the overall system performance and identify bottleneck agents?} Addressing this challenge is critical for diagnosing and improving multi-agent LLM systems \citep{liu2023agentbench, liu2023dynamic}.

To address this challenge, recent works adopt methods based on cooperative game theory, which measure marginal contributions by re-evaluating the MAS under different agent coalitions---subsets formed by sampled agents. For example, \citet{cui2025efficient} use leave-one-out (LOO) attribution, while others \citep{xia2026hivemind, ebrahimi2025adversary} estimate Shapley values \citep{shapley1953value, lundberg2017unified} over multiple coalitions. However, evaluating many coalitions can be computationally expensive, and ablating structurally essential agents, such as orchestrators, can make the system non-executable. Existing methods address these issues in different ways, including replacing agent ablation with in-context simulation \citep{cui2025efficient}, replacing an agent's backbone model rather than removing it \citep{yang2025s}, or restricting coalitions according to system topology \citep{xia2026hivemind}. Yet it remains unclear how these choices---which coalitions are considered and how agents are intervened on---shape attribution across MAS structures, leaving the trade-offs among attribution methods and their suitability for different interventions poorly understood.

% Despite progress, how these intervention protocols and choices of the coalition distribution interact across collaboration topologies has not been thoroughly investigated. As a result, the trade-offs among design choices for agent attribution remain poorly understood, leaving practitioners without clear guidance on how to select attribution methods for different MAS.

\begin{figure}[h!]
    \centering
    \includegraphics[width=0.8\linewidth]{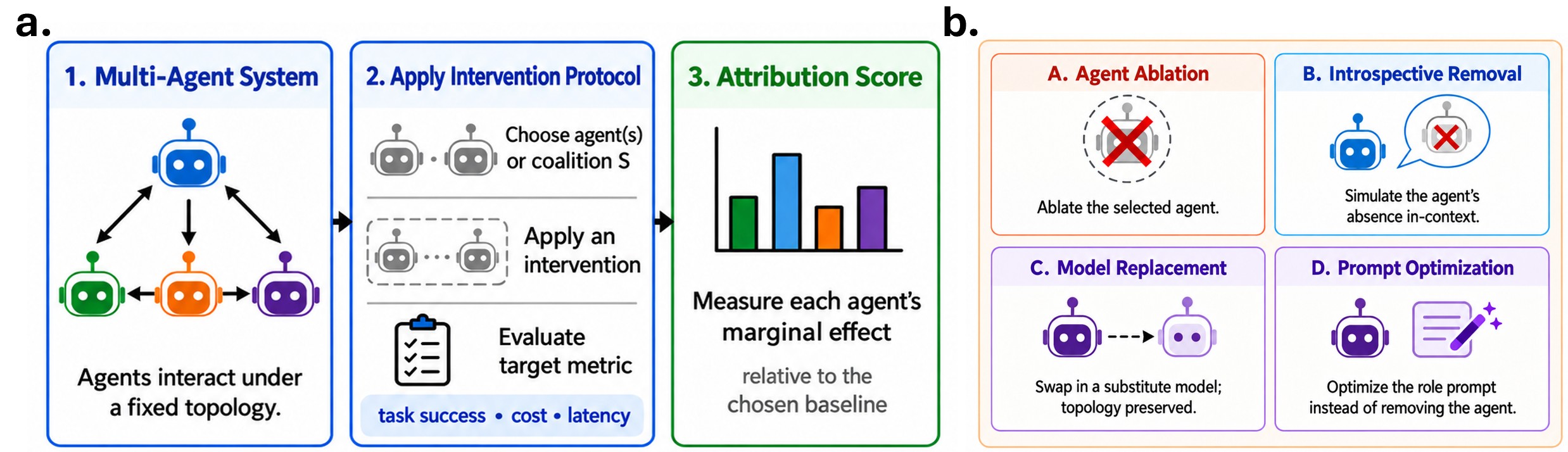}
    \caption{\textbf{(a)} Overview of intervention-based attribution for MAS and \textbf{(b)} intervention protocols.}
    \label{fig:concept_figure}
    \vspace{-1em}
\end{figure}

To enable this investigation, we formalize agent attribution as an \emph{intervention-conditioned cooperative game} \citep{covert2021explaining}. Under this formulation, an attribution query is parameterized by three variables: (i) the intervention protocol, (ii) the coalition distribution, and (iii) the target metric. Attribution scores are therefore defined relative to an evaluation query, rather than as a universal importance score. Using this framework, our analysis reveals several key insights. First, distinct intervention approaches induce different attribution games: Agent ablation concentrates attribution on structurally indispensable roles, e.g., orchestrator, and introspective removal with an LLM judge is not faithful to this ablation. Second, despite its simplicity, LOO effectively identifies key agents while reducing computational costs by $2.2\times\text{--}11.9\times$ compared to methods that require combinatorial coalitions across datasets. Under our framework, prompt optimization defines a distinct intervention-conditioned attribution game. Using prompt-optimization LOO, we identify effective optimization targets while reducing attribution cost by \(12\%\) relative to prior work \citep{xia2026hivemind} and using \(79\%\) fewer tokens than prompt-optimization Shapley. Finally, we show that agent importance is objective-dependent. When auditing a medical MAS, we find that agent contributions to diagnostic accuracy and ethical alignment are often decoupled; agents that improve accuracy may actively degrade ethics scores. Overall, we establish a unified framework for agent attribution, providing a principled and practical basis for selecting attribution methods and guiding MAS interventions.

\vspace{-0.5em}
\section{Related Works}
\noindent\textbf{Instance-level attribution and trace-based diagnosis.}
\citet{liu2023dynamic} propagate peer ratings through communication
graphs, while \citet{cui2025efficient} approximate leave-one-out removal through in-context simulation. Failure-diagnosis methods localize errors to agents and steps \citep{zhang2025agent} or characterize recurring interaction failures \citep{cemri2026multi}. DoVer \citep{ma2026dover} goes beyond trace inspection by testing message and plan interventions through re-execution. Our focus is complementary to instance-specific repair: estimating intervention effects across a task distribution to guide system-level intervention.

\noindent\textbf{Credit assignment for agent coordination.}
Beyond post-hoc analysis, contribution estimates can also shape agent interactions during execution. \citet{ebrahimi2025adversary} dynamically update agent credibility for adversary-resistant aggregation, while Shapley-Coop \citep{hua2026shapley} uses Shapley-based credit assignment for negotiation and reward redistribution. These methods use attribution to coordinate agent behavior during execution; we instead study how the underlying game formulation shapes the distribution of attribution values.

\noindent\textbf{Global agent attribution and optimization.}
At the system level, attribution has also been used to guide architecture and optimization decisions. \citet{yang2025s} evaluate functional components through Shapley values under backbone-model substitution. \citet{xia2026hivemind} introduce CG-OPO to account for workflow dependencies, assigning zero utility to non-viable coalitions, and use removal-based attribution to select agents for prompt refinement. Taken together, however, existing approaches typically instantiate attribution with fixed design choices, making it difficult to isolate how these choices shape the resulting attribution. We address this gap with three contributions: we (i) unify previously separate design choices within a common attribution framework, (ii) systematically benchmark their effects across diverse MAS structures, and (iii) connect attribution to downstream system analysis and intervention.

\vspace{-0.5em}
\section{A Unified Framework for Agent Attribution}
In this work, we introduce a unified framework for agent attribution by formalizing it as a {protocol-conditioned cooperative game} \citep{covert2021explaining}. This view unifies existing approaches and makes explicit what question each attribution design answers.

\subsection{Agent Attribution as a Cooperative Game}
We formalize agent attribution through intervention games that quantify how changes to individual agents affect system-level behavior. This establishes the foundation of our work.
\begin{definition}
    A multi-agent system induces a baseline game represented by the tuple \((\mathcal A,\Gamma,\nu)\), where \(\mathcal{A}=\{a_1,\dots,a_n\}\) is the set of agents, \(\Gamma\) specifies the reference MAS architecture, including roles, routing, communication structure, and execution rules for configurations with absent agents, and \(\nu:2^{\mathcal A}\rightarrow\mathbb{R}\) is the characteristic utility function. Given a task distribution \(\mathcal D\), the utility of an active coalition \(C\subseteq\mathcal A\) is
    \begin{equation}
    \nu(C)
    =
    \mathbb{E}_{x\sim\mathcal{D},\,\xi\sim\Xi}
    \left[
    \mathcal{G}\left(\tau_{C,x,\xi}^{\Gamma}\right)
    \right].
    \end{equation}
    where \(\tau_{C,x,\xi}^{\Gamma}\) denotes the execution trace generated by the original agents in \(C\), with all other agents absent, on input \(x\) under architecture \(\Gamma\) and randomness \(\xi\). Here, \(\xi\) captures stochasticity from the LLM and/or environment dynamics, and \(\mathcal G\) maps the execution trace to a scalar score, such as task success or token usage.
\end{definition}
% We extend this baseline with protocol-specific reference states $\mathbf b=(b_1,\ldots,b_n)$, such as model replacement or agent ablation. For each intervention target \(a_i\), \(i\in[n]\), we define the state of every agent \(a_k\), \(k\in[n]\), under coalition \(S\) as
% \begin{equation}
%     m_k^{i,\mathbf b}(S)
%     =
%     \begin{cases}
%         b_i, & k=i \text{ and } a_i\in S,\\
%         \varnothing, & k\neq i \text{ and } a_k\notin S,\\
%         a_k, & \text{otherwise}.
%     \end{cases}
% \end{equation}
We generalize this baseline to intervention protocols by introducing protocol-specific reference states, denoted by $\mathbf{b} = (b_1, \dots, b_n)$. For an intervention coalition $S \subseteq \mathcal{A}$, agents in $S$ adopt their respective reference states, while all other agents retain their original implementations:
\begin{equation}
m_i^{\mathbf{b}}(S) = 
\begin{cases}
b_i, & \text{if } a_i \in S, \\
a_i, & \text{if } a_i \notin S.
\end{cases}
\end{equation}
We define the corresponding expected utility as:
\begin{equation}
\nu_{\mathbf{b}}(S) = \mathbb{E}_{x \sim \mathcal{D}, \xi \sim \Xi} \left[ \mathcal{G} \left( \tau^{\Gamma}(\mathbf{m}^{\mathbf{b}}(S), x, \xi) \right) \right].
\end{equation}
For any $S \subseteq \mathcal{A} \setminus \{a_i\}$, the difference $\nu_{\mathbf{b}}(S \cup \{a_i\}) - \nu_{\mathbf{b}}(S)$ captures the marginal effect of intervening on $a_i$, holding the intervention status of all other agents fixed. We therefore define:
\begin{equation}\label{eq:intervention_game}
\phi_i^{\mathbf{b}} = \sum_{S \subseteq \mathcal{A} \setminus \{a_i\}} w_i(S) \left[ \nu_{\mathbf{b}}(S \cup \{a_i\}) - \nu_{\mathbf{b}}(S) \right].
\end{equation}
Thus, $\phi_i^{\mathbf{b}}$ quantifies the effect of intervening on $a_i$, aggregated across intervention coalitions according to $w_i$. An attribution query is therefore specified by $(\mathcal{G}, w_i, \mathbf{b})$, where $\mathcal{G}$ specifies the target behavior, $w_i$ determines the weighting over intervention coalitions, and $\mathbf{b}$ defines the intervention protocol.

\subsection{Coalition Distributions for MAS}
Rather than treating the coalition distribution $w_i(S)$ as universally fixed, our framework treats it as a design choice. The weights $w_i(S)$ specify which coalitions are used to estimate the contribution of agent $a_i$, and different choices correspond to different attribution methods and computational costs.

We consider a spectrum of coalition distributions. Leave-One-Out (LOO) measures the effect of intervening on \(a_i\) while all other agents remain unchanged, making it efficient but less sensitive to interactions that emerge when other agents are also intervened on. In contrast, combinatorial methods are more computationally expensive but evaluate marginal contributions across coalitions of varying sizes. The Shapley value \citep{shapley1953value} averages over all possible coalitions, while Owen values \citep{owen1977essays} incorporate group structure by partitioning agents into modules. Myerson values \citep{myerson1977graphs} can additionally impose graph-based constraints through connected components of an undirected attribution graph derived from the MAS communication structure. We compare these choices to examine how coalition distributions affect attribution across MAS architectures. Further details are provided in \Cref{apx:attribution_methods}.

\subsection{Intervention Protocols in MAS}\label{sec:removal_protocols}
The reference vector $\mathbf b$ determines the intervention evaluated by attribution. We consider four protocols: agent ablation, introspective removal, model replacement, and prompt optimization. Under our convention, attribution measures the utility change from applying the intervention to $a_i$.
\vspace{0.2em}

\noindent\textbf{Agent ablation.}
Agent ablation sets $b_i=\varnothing$, removing $a_i$ from the evaluated configuration. While this aligns with the notion of player removal in classical cooperative game theory, ablating indispensable roles such as routers or orchestrators can result in system failures. A common workaround is to assign zero utility to non-executable coalitions \citep{xia2026hivemind}, but it may conflate an agent's task contribution with its role in maintaining system executability.
\vspace{0.2em}

\noindent\textbf{Introspective removal.} 
\citet{cui2025efficient} approximate ablation by prompting agents to revise their answers while disregarding another
agent's responses. We denote simulated absence by $b_i=\varnothing_{\mathrm{sim}}$, using this notation as shorthand for counterfactual utility estimation rather than a deployable implementation. The utility is estimated through in-context reasoning or an LLM judge instead of executing the ablated configuration. This approximates agent ablation, with fidelity depending on the evaluator and its prompts.
\vspace{0.2em}

\noindent\textbf{Model replacement.}
Prior work evaluates model replacement in workflows with four predefined components \citep{yang2025s}. We study this intervention across MAS with different agent roles and communication topologies. Specifically, we set $b_i=\tilde a_i$, where $\tilde a_i$ uses a different backbone model while retaining the agent's role,
prompt, and tool access. Unlike ablation, replacement preserves
the role and configured communication graph within each comparison. The resulting attribution also depends on the chosen replacement model, measuring the utility change from model substitution.
\vspace{0.2em}

\noindent\textbf{Prompt optimization.}
Prior work \citep{xia2026hivemind} uses removal-based Shapley
attribution to select agents for prompt refinement. In contrast,
we treat prompt optimization \citep{khattab2024dspy,agrawal2026gepa}
itself as an intervention when computing attributions. We set \(b_i=a_i^{\mathrm{opt}}\), where \(a_i^{\mathrm{opt}}\) uses an optimized prompt \(\tilde p_i\) while retaining the agent's role, backbone model, tool access, and configured communication links. With $\tilde p_i$ held fixed across coalition evaluations, the resulting attribution measures the utility change from that update.
% , enabling target selection based on update-specific gains.
\vspace{-0.5em}

% While agent ablation and introspective removal both define the attribution protocol through the absence of agent \(a_i\), our framework broadens this view: A removal protocol needs not eliminate the agent itself. We introduce \emph{model replacement}, a topology-preserving protocol that substitutes an agent's backbone model while keeping its role instantiated. Formally, for agent \(a_i\), we define a replacement protocol \(b_i=\tilde{a}_i\), where \(\tilde{a}_i\) may be weaker, comparable, or stronger than \(a_i\) under a predefined capability criterion, such as model scale or tool access. Because every role remains instantiated, the MAS communication graph is preserved. The resulting attribution measures the marginal utility of changing the model assigned to an agent: A weaker replacement estimates the contribution of the original model's additional capacity, while a stronger replacement estimates the potential benefit of upgrading that role. This enables capability-allocation analysis: informing agent \emph{harness engineering} by identifying which roles require stronger backbone models, additional tools, or cheaper substitutes \citep{meng2026agent}. 

\subsection{Behavior Metrics}\label{sec:behavior_metrics}
The final query component, $\mathcal{G}$, specifies what behavior is being attributed. Given an execution trace, it returns a scalar score, such as task success, token usage, or failure rate. The coalition utility $\nu_\mathbf{b}(S)$ is the expected value of this score over inputs and system randomness. Therefore, the choice of $\mathcal{G}$ determines how attribution scores should be interpreted: Accuracy metrics attribute contributions to task performance, while cost-based metrics attribute contributions to resource use.

Together, these components formalize MAS attribution in terms of the intervention protocol $\mathbf{b}$, the attribution distribution $w_i(S)$, and the behavior metric $\mathcal{G}$, providing a unified framework for analyzing MAS under different attribution queries.
\label{sec:method}

\vspace{-0.5em}
\section{Benchmarks and MAS Architectures}
\vspace{-0.5em}

% We now describe the experimental environment used to instantiate our framework and study agent attribution methods.

\subsection{Benchmark Datasets}
We instantiate our environments using the framework of \citet{kim2025towards}\footnote{\url{https://github.com/ybkim95/agent-scaling}} and evaluate on four benchmarks: (1) PlanCraft \citep{dagan2024plancraft}, a Minecraft-based environment for assessing planning and tool use; (2) WorkBench \citep{styles2024workbench}, which evaluates planning and tool selection in business workflows; (3) BrowseComp-Plus \citep{chen2025browsecomp}, which evaluates deep-research agents by retrieval, reasoning, and tool usage over human-verified documents; and (4) LiveCodeBench \citep{jain2025livecodebench}, which evaluates code generation on competitive-programming problems. Please refer to \Cref{apx:datasets} for dataset details.

\subsection{Communication Topologies} 
\begin{figure}[h!]
    \centering
    \includegraphics[width=1.0\linewidth]{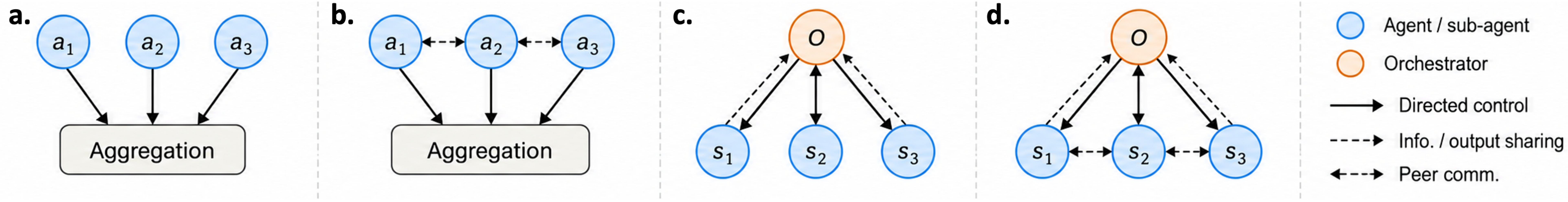}
    \caption{MAS communication graphs: \textbf{(a)} Independent \textbf{(b)} Decentralized \textbf{(c)} Centralized \textbf{(d)} Hybrid. }
    \label{fig:mas_topologies}
    \vspace{-1em}
\end{figure}

To study MAS behavior across these environments, we adopt four common collaboration topologies from \citet{kim2025towards}. Each topology is defined by a fixed set of agents \(\mathcal{A}\) and directed communication links \(E\) (\Cref{fig:mas_topologies}).

\noindent\textbf{Independent:} $E_{\mathrm{ind}}=\varnothing$, where agents operate independently without inter-agent communication. The system returns the first successful attempt, or the first completed attempt if none succeeds.

\noindent\textbf{Decentralized:} $E_{\mathrm{decent}} = \{(a_i, a_j): a_i,a_j \in \mathcal{A}, i\neq j\}$, a peer-to-peer network where agents work in parallel and share their findings with all peers between rounds.

\noindent\textbf{Centralized:} $E_{\mathrm{cent}}
= \{(a_{\mathrm{orch}},a_i),(a_i,a_{\mathrm{orch}})
: a_i\in\mathcal{A}\setminus\{a_{\mathrm{orch}}\}\}$, where an orchestrator \(a_{\mathrm{orch}}\) assigns subtasks, exchanges findings with each worker over multiple rounds, and synthesizes the final answer.

% \noindent\textbf{Hybrid:} $E_{\mathrm{hyb}} = E_{\mathrm{cent}} \cup E_{\mathrm{peer}}$, which augments the centralized topology with peer-to-peer links $E_{\mathrm{peer}} = \{(a_i, a_j): a_i, a_j \in \mathcal{A}\setminus\{a_{\mathrm{orch}}\},\, i\neq j\}$ connecting all pairs of workers.
\noindent\textbf{Hybrid:}
$E_{\mathrm{hyb}}=E_{\mathrm{cent}} \cup\{(a_i,a_j): a_i,a_j\in\mathcal{A}\setminus\{a_{\mathrm{orch}}\},\, i\neq j\}$,
which augments the centralized topology with peer-to-peer links between all workers.

\subsection{MAS Configurations}\label{sec:mas_configs}
For each benchmark, we use a fixed set of role-specialized agents with distinct personas, task-specific prompts, and tool access. To keep exact computation of combinatorial attribution methods, such as Shapley values, feasible, we use five agents for PlanCraft and WorkBench and four subagents for LiveCodeBench and BrowseComp-Plus; centralized and hybrid topologies include one additional orchestrator. Following \citet{kim2025towards}, we report mean task success. For our primary experiments on four datasets, all agents use GPT-5.4 with medium reasoning, and communicating topologies run for at most five rounds. We additionally evaluate alternative backbones in \Cref{sec:alternative_backbones}. Detailed agent roles, implementation details, and prompt templates are provided in \Cref{apx:agent_configs,apx:prompt_template}.

% For each benchmark, we employ a fixed set of role-specialized agents—each defined by a distinct persona, task-specific prompt, and tool access. To enable exact attribution computation while keeping combinatorial methods tractable, we instantiate PlanCraft and WorkBench with five subagents each, powered by Gemini-2.5-Flash \citep{comanici2025gemini} and Claude 3.5 Haiku \citep{anthropic2024model}, respectively. BrowseComp-Plus uses four subagents powered by GPT-5-mini \citep{singh2025openai}. LiveCodeBench uses four subagents powered by GPT 5.4. In the centralized and hybrid topologies, we add a single orchestrator agent; no orchestrator is used in the independent or decentralized settings. Detailed agent roles, implementation details, and prompt templates are provided in \Cref{apx:agent_configs,apx:prompt_template}.
\vspace{-0.5em}
\subsection{Attribution Computation \& Intervention Protocol Implementation}
\label{sec:removal_implementation}
\vspace{-0.5em}

For agent ablation, we remove the target agent from the system. For introspective removal, an LLM judge estimates the effect of removing the target agent without modifying the deployed MAS, using the prompt in \Cref{prompt:judge_prompt}. For model replacement, we substitute the target agent's backbone with GPT-OSS-120B \citep{agarwal2025gpt}, the strongest open-weight model from the same provider. For prompt optimization, we run GEPA \citep{agrawal2026gepa} for five iterations with GPT-5.5 as the reflector, jointly optimizing all agent prompts on 100 examples disjoint from evaluation and split evenly between training and validation. Optimization uses the benchmark-specific evaluation objective, and each resulting agent prompt defines its intervention. 

For each benchmark, we use 100 instances for attribution estimation and a disjoint set of 200 instances for evaluating top-\(k\) interventions, with agents ranked in descending order of their protocol-specific attribution. Each MAS configuration is evaluated over three independent runs.

\vspace{-0.5em}
\section{Experiments \& Results}
\vspace{-0.5em}
We organize our experiments around practical questions and assumptions that arise in prior work. For each, we use our framework to isolate a relevant attribution design choice—intervention protocol, coalition distribution, or evaluation objective—and examine how it shapes agent importance across benchmarks and communication topologies.

\vspace{-0.5em}
\subsection{Effect of Intervention Protocols}
\vspace{-0.5em}

\noindent \textbf{Fidelity of introspective removal.}
% We first examine introspective removal \citep{cui2025efficient}, where an independent LLM judge (\Cref{prompt:judge_prompt}) is provided with agent conversation to estimate coalition (subset) values without rerunning ablations. To assess its fidelity, we compare its coalition utilities against agent ablation and compute $R^2$ and Spearman's $\rho$. 
To avoid costly re-execution, prior work \citep{cui2025efficient} derives attribution via counterfactual reasoning about MAS behavior when a targeted agent is removed. This approximation is reliable only if the inferred counterfactual utilities match those obtained by agent ablation. To ensure our analysis remains agnostic to the choice of coalition distribution, we directly compare introspective removal and agent ablation utilities across all $2^n-1$ coalitions and compute $R^2$ and Spearman's $\rho$.

\begin{figure}[t]
    \centering
        \begin{minipage}[c]{0.48\linewidth}
        \centering
        \includegraphics[width=\linewidth]{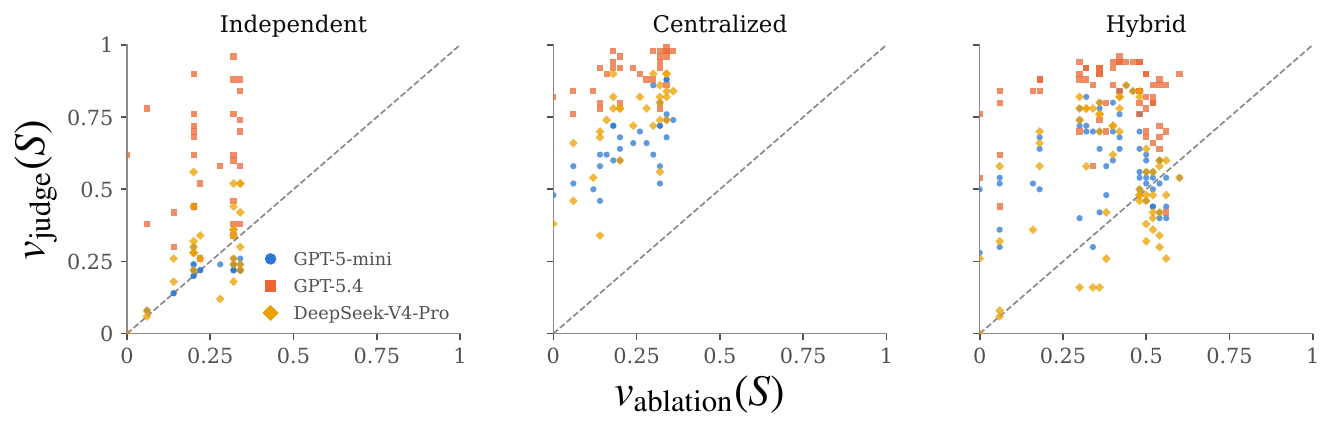}
        \vspace{-2em}
        \captionof{figure}{Introspective removal ($v_{\text{Judge}}$, $y$-axis) versus agent ablation coalition utility ($v_{\text{ablation}}$, $x$-axis) on PlanCraft.}
        \label{fig:judge_pr_plancraft}
    \end{minipage}
    \hfill
    \begin{minipage}[c]{0.5\linewidth}
        \centering
        \vspace{-0.5em}
        \captionof{table}{Agreement ($R^{2}$ / Spearman $\rho$) between introspective removal and agent-ablation coalition values on PlanCraft across all coalitions and backbone models, averaged over three runs.}
        \resizebox{\linewidth}{!}{%
        \begin{tabular}{lcccc}
        \toprule
        \textbf{Judge Model} &
        \textbf{Ind.} &
        \textbf{Decent.} &
        \textbf{Cent.} &
        \textbf{Hybrid} \\
        \midrule
        DeepSeek-V4-Flash  & $-1.05$ / 0.47 & $-2.03$ / 0.52 & \textbf{-11.93} / \textbf{0.80} & \textbf{-0.99} / \textbf{0.14} \\
        DeepSeek-V4-Pro    & $-0.77$ / 0.43 & $-1.12$ / 0.56 & $-21.46$ / 0.69 & $-2.16$ / 0.01 \\
        GPT-5-mini         & \textbf{0.38} / \textbf{0.47} & \textbf{0.78} / \textbf{0.72} & $-15.61$ / 0.74 & $-2.16$ / $-0.08$ \\
        GPT-5.4            & $-17.90$ / 0.18 & $-10.21$ / 0.56 & $-39.48$ / 0.67 & $-7.79$ / $-0.06$ \\
        % DeepSeek-V4-Flash
        %     & 0.18 / 0.60
        %     & \textbf{0.18} / 0.49
        %     & \textbf{-1.14} / 0.56
        %     & -2.94 / 0.54 \\
        % DeepSeek-V4-Pro
        %     & -0.77 / 0.43
        %     & -1.12 / \textbf{0.56}
        %     & -21.46 / 0.69
        %     & \textbf{-2.16} / 0.01 \\
        % GPT-5-mini
        %     & \textbf{0.42} / \textbf{0.63}
        %     & 0.15 / 0.55
        %     & -9.63 / \textbf{0.81}
        %     & -9.94 / \textbf{0.65} \\
        % GPT-5.4
        %     & -17.90 / 0.18 
        %     & -10.21 / \textbf{0.56}
        %     & -39.48 / 0.67
        %     & -7.79 / -0.06 \\
        \bottomrule
        \end{tabular}%
        }
        \label{tab:coalition-agreement-plancraft}
    \end{minipage}
    \vspace{-0.5em}
\end{figure}

As shown in \Cref{fig:judge_pr_plancraft}, introspective removal generally overestimates coalition utility relative to agent ablation. At the instance level, introspective predictions disagree with actual execution on \(30\text{--}52\%\) of PlanCraft cases, with errors dominated by false positives: they often predict success when the instance actually fails. Moreover, increasing model capability does not consistently improve fidelity; GPT-5.4 does not outperform GPT-5-mini and exhibits stronger overestimation. This bias is reflected in coalition-level agreement (\Cref{tab:coalition-agreement-plancraft}). In the independent topology, the strongest judge (GPT-5-mini) achieves moderate agreement \(R^2=0.38; \rho=0.47, p=.008\), while all judge models obtain negative \(R^2\) in centralized topologies despite moderate rank correlation. Agreement is even weaker on a more challenging BrowseComp-Plus (\Cref{tab:coalition-agreement-combined}). We additionally test standard judge-enhancement strategies \citep{wang2022self,liu2023g,madaan2023self}, but none consistently improves agreement (\Cref{tab:judge_design_sensitivity}). Taken together, these results show that introspective removal, despite being substantially cheaper (\Cref{tab:attribution-distribution}), does not reliably approximate agent ablation, and that stronger judges or refinement strategies do not close this gap.

\vspace{0.5em}
\vspace{-0.5em}
\begin{figure}[t]
    \centering
    % Left: attribution heatmap
    \begin{minipage}[t]{0.48\linewidth}
        \vspace{0pt}
        \centering
        \includegraphics[width=\linewidth]{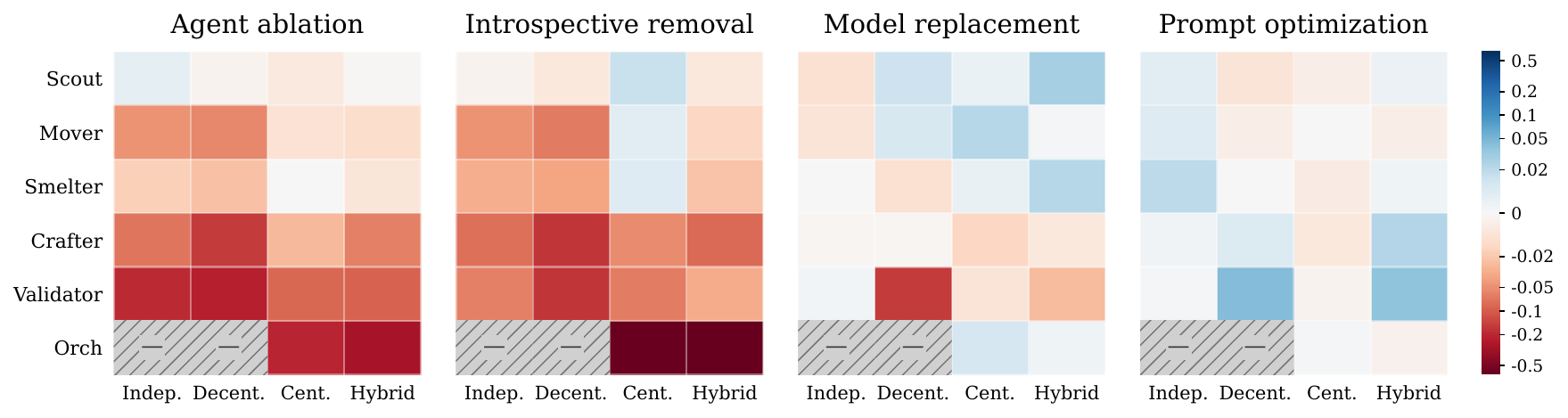}
        \caption{Agent Shapley values across intervention protocols on PlanCraft. Blue (red) indicates positive (negative) effects on task success. }
        \label{fig:attribution_dis_plancraft}
    \end{minipage}
    \hfill
    % Right: entropy + token table
    \vspace{-0.5em}
    \begin{minipage}[t]{0.5\linewidth}
        \centering
        \scriptsize
        \captionof{table}{ Normalized attribution entropy and token usage averaged across PlanCraft. Lower entropy indicates more concentrated attribution; shaded cells denote entropy $<0.6$. }
        \label{tab:attribution_distribution_avg}
        \vspace{-0.5em}
        \resizebox{\linewidth}{!}{%
        \begin{tabular}{lrrrr|r}
        \toprule
        \textbf{Intervention}
        & \textbf{Ind.}
        & \textbf{Decent.}
        & \textbf{Cent.}
        & \textbf{Hybrid}
        & \textbf{Tokens} \\
        \midrule
        Agent ablation        & 0.724 & 0.713 & \cellcolor{gray!25}0.580 & \cellcolor{gray!25}0.599 & 2.17M  \\
        Introspective removal & 0.860 & 0.808 & \cellcolor{gray!25}0.422 & \cellcolor{gray!25}0.424 & 409.7K  \\
        Model replacement     & 0.711 & \cellcolor{gray!25}0.469 & 0.929 & 0.817 & 1.51M  \\
        Prompt optimization   & 0.796 & 0.657 & 0.811 & 0.728 & 8.13M \\
        \bottomrule
        \end{tabular}
        }
    \end{minipage}
    \vspace{-1em}
\end{figure}

\noindent \textbf{Distinct notions of agent importance.} The disagreement between introspective removal and actual ablation suggests that agent importance is intervention-dependent rather than intrinsic. While prior work typically evaluates contribution under a single fixed intervention  \citep{cui2025efficient,hua2026shapley,xia2026hivemind}, our framework makes this choice explicit: each intervention induces a distinct attribution game (\Cref{fig:attribution_dis_plancraft},\Cref{sec:removal_protocols}). To quantify this, we compare the concentration of attributions\footnote{Shapley values are shown here; results for other coalition distributions are reported in \Cref{tab:attribution-distribution-loo,tab:attribution-distribution-owen,tab:attribution-distribution-myerson}.} across intervention games using normalized Shannon entropy \citep{shannon1948mathematical}\footnote{$H_{\mathrm{norm}}(\phi) = -(\log n)^{-1}\sum_i p_i \log p_i$, where $p_i$ is the normalized absolute attribution of agent $i$.}.

As shown in \Cref{tab:attribution_distribution_avg}, intervention protocols differ most clearly under hierarchical coordination on PlanCraft. In centralized and hybrid topologies, agent ablation and introspective removal yield concentrated attribution (\(H_{\mathrm{norm}}=0.580/0.599\) and \(0.422/0.424\), respectively), whereas model replacement (\(0.929/0.817\)) and prompt optimization (\(0.811/0.728\)) distribute attribution more broadly. This contrast reflects the intervention being measured: removal-based protocols emphasize agents whose absence disrupts the system, whereas replacement and prompt optimization measure the value of modifying agents while preserving their roles and communication structure. A broadly similar pattern holds under LOO, Owen, and Myerson (\Cref{tab:attribution-distribution-loo,tab:attribution-distribution-owen,tab:attribution-distribution-myerson}). The protocols also differ in attribution cost, with prompt optimization requiring additional optimization and MAS execution, while introspective removal relies only on judge-based evaluation (\Cref{tab:attribution-distribution}). 

\vspace{-0.2em}
\begin{tcolorbox}[colback=blue!5!white,    top=0.25em,
    bottom=0.25em,  colframe=blue!75!black]
\noindent \textbf{Key Takeaway:} Agent importance is not a single fixed quantity: importance depends on the intervention used to define it, leading to distinct attribution games and trade-offs.
\end{tcolorbox}
\vspace{-0.8em}

\subsection{Choice of Coalition Distribution}
\vspace{-0.5em}
% \begin{tcolorbox}[colback=blue!5!white, colframe=blue!75!black]
% \textbf{Key Takeaway:} LOO can match the intervention
% performance of methods that evaluate combinatorial coalitions while incurring substantially lower cost
% \end{tcolorbox}

\noindent\textbf{Rank agreement across coalition distributions.}
Existing work typically adopts a single coalition distribution, such as Shapley, without examining whether this choice changes which agents are identified as important \citep{cui2025efficient, ebrahimi2025adversary, yang2025s, xia2026hivemind}. We compare LOO, Shapley, Owen, and Myerson (\Cref{apx:attribution_methods}) and measure rank agreement using Spearman correlation. 

\begin{figure}[t!]
    \centering
    \includegraphics[width=\linewidth]{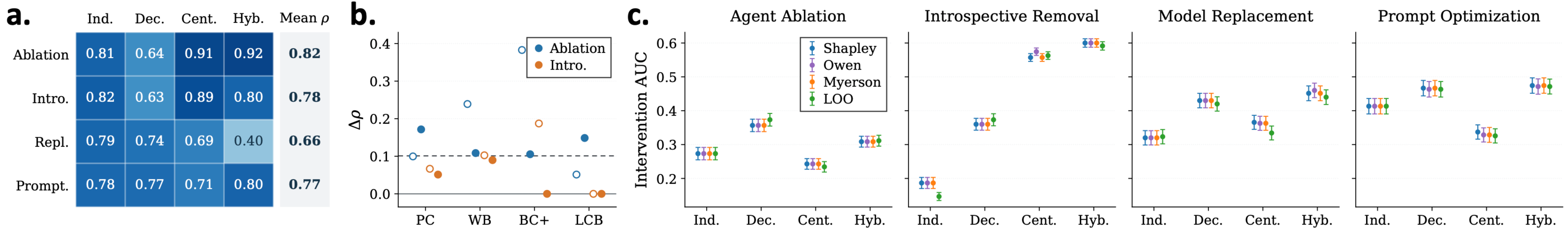}
    \caption{ \textbf{(a)} LOO--Shapley Spearman correlation across intervention protocols and communication topologies, averaged over datasets. \textbf{(b)} Change in correlation by dataset. Hollow/solid markers denote centralized/hybrid topologies. \textbf{(c)} Rank-guided intervention AUC with 95\% confidence intervals across coalition distributions and communication topologies on PlanCraft. }
    \label{fig:auc_plancraft}
    \vspace{-2em}
\end{figure}

Across intervention protocols and communication topologies, the combinatorial methods generally yield highly consistent rankings despite differences in attribution magnitude, with pairwise $\rho=0.76\text{--}1.0$ among Shapley, Owen, and Myerson, averaged over datasets and interventions (\Cref{fig:rank_agreement_methods}). Notably, LOO also shows agreement with Shapley, with $\rho=0.66\text{--}0.82$ across interventions, averaged over topologies (\Cref{fig:auc_plancraft}(a)). To better understand this agreement, we further examine how it varies across intervention protocols and system structures. For agent ablation, we find more stable rankings when the system contains an essential agent, such as the Coder in LiveCodeBench or the orchestrator in centralized and hybrid systems, where coalitions without that agent have low or zero utility. 

% These constraints concentrate Shapley contributions on coalitions containing the essential agent, shifting Shapley toward larger coalitions and closer to the full-coalition setting evaluated by LOO. 
To quantify the effect of the essential agent, we recompute Shapley values with the essential agent always present.\footnote{For essential agent \(e\), we hold \(e\) un-intervened and recompute Shapley values over intervention coalitions \(S\subseteq\mathcal A\setminus\{e\}\)} Removing this essential-agent-induced weighting reduces LOO--Shapley correlation among the remaining agents by a mean of \(0.113\) (\Cref{fig:auc_plancraft}(b)). Similarly, model replacement also weakens these structural utility gaps by preserving agent roles and yields lower rank agreement (\(\rho=0.66\); \Cref{fig:auc_plancraft}(a)). A complementary case is agent redundancy: an agent may be redundant in the full team yet useful in smaller coalitions, causing LOO and Shapley to diverge. For example, on WorkBench, two agents share the same CRM--Email Generalist role (\Cref{sec:agent-cfg-workbench}), and adding the second provides no additional gain. LOO therefore assigns near-zero importance to most agents when evaluated against the full coalition, whereas Shapley captures their contributions across smaller coalitions, yielding much lower agreement (\(\rho=0.10\text{--}0.50\); \Cref{fig:rank_agreement_methods}). 

\textbf{Intervention effectiveness.}
Finally, we evaluate whether ranking agreement translates into similar intervention outcomes by sequentially intervening on agents according to their attribution ranking. We summarize effectiveness using top-\(k\) intervention AUC with 95\% confidence intervals \citep{petsiuk2018rise}. Within each intervention protocol and communication topology, rank-guided intervention AUCs differ marginally across coalition distributions (\Cref{fig:auc_plancraft}(c)). This holds not only for the independent topology but also for systems with communication graphs. For example, under decentralized agent ablation, top-\(k\) AUCs range from \(0.263\,(0.026)\) to \(0.270\,(0.033)\). We observe the same pattern on WorkBench, BrowseComp-Plus, and LiveCodeBench (\Cref{fig:auc_workbench_browsecomp_livecodebench}); paired \(\Delta\)AUC comparisons across coalition distributions further confirm that these differences remain small (\Cref{fig:auc_paired_diff_gpt54}).

\vspace{-0.2em}
\begin{tcolorbox}[colback=blue!5!white,    top=0.25em,
    bottom=0.25em, colframe=blue!75!black]
\noindent \textbf{Key takeaway:} When the attribution game induces clear utility gaps—e.g., via intervention protocols or agent capability asymmetries—LOO can serve as an effective proxy for combinatorial attribution, recovering similar rankings at \(2.2\text{--}11.9\times\) lower token cost (\Cref{tab:cost_tokens_merged}).
\end{tcolorbox}
\vspace{-0.8em}

\subsection{Performance--Cost Trade-offs of Intervention Protocols}

% \begin{tcolorbox}[colback=blue!5!white, colframe=blue!75!black]
% \textbf{Key Takeaway:} Attribution can identify effective intervention targets, but the resulting cost--performance trade-off depends on how agents are intervened on.
% \end{tcolorbox}

% \noindent Prior work \citep{xia2026hivemind,ebrahimi2025adversary,yang2025s} uses attribution to select intervention targets but does not systematically compare performance--cost trade-offs across interventions. We evaluate these trade-offs by ranking agents with LOO under each intervention protocol, applying the intervention to the selected \(k\) agents, and measuring task success and inference token usage.

\noindent Prior work \citep{xia2026hivemind,ebrahimi2025adversary,yang2025s} uses attribution to select intervention targets but overlooks the resulting performance--cost trade-offs. This is particularly important in MASs, where inference costs can be substantial \citep{smit2023should,zeng2025s2}. We therefore evaluate these trade-offs by ranking agents with LOO, applying the corresponding intervention to the selected \(k\) agents, and measuring task success and inference token usage.

\begin{figure}[t!]
    \centering
    \includegraphics[width=.9\linewidth]{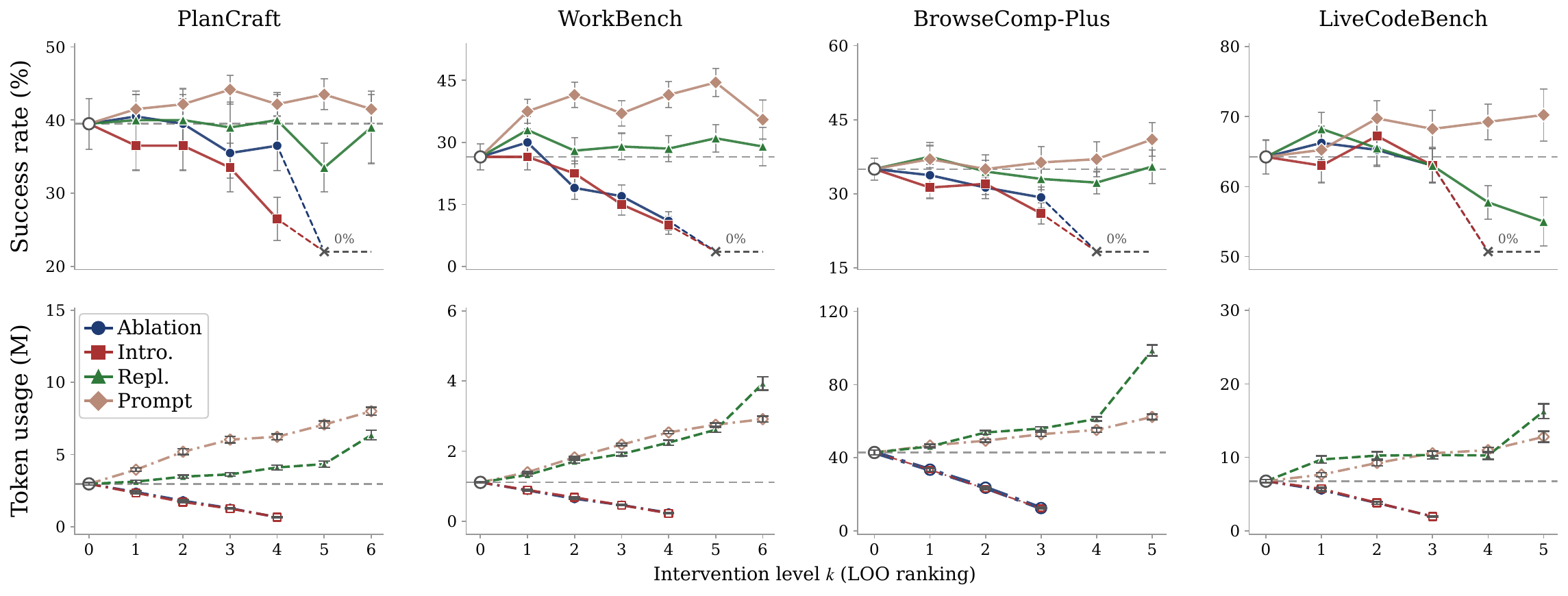}
    \caption{Task success (top) and inference cost (bottom) under top-\(k\) agent interventions across datasets. Error bars show 95\% confidence intervals over communication topologies and three runs.}
    \label{fig:cost_reduction_plancraft}
    \vspace{-1.5em}
\end{figure}

\noindent\textbf{Agent ablation and introspective removal.} Our results show that removing agents with the highest ablation attribution can improve both performance and inference cost, but these gains diminish with deeper removal (\Cref{fig:cost_reduction_plancraft}). For example, on WorkBench, ablation at \(k=1\) improves success by 3.5 percentage points while reducing token usage by 26.5\% relative to the full system. However, on the more challenging BrowseComp-Plus benchmark, performance declines from \(k=1\). Overall, further removal reduces token usage but degrades performance significantly across datasets. For introspective removal, agents are removed based on judge-based attribution ranking. This yields similar token savings but less consistent performance than direct ablation.

\noindent\textbf{Model replacement.}
Model replacement preserves agent roles while changing their underlying models. With GPT-OSS-120B, replacing the \(k=1\text{--}3\) highest-ranked agents on WorkBench improves success by \(1.5\text{--}6.5\) percentage points over the baseline, while increasing total token usage by \(19\text{--}73\%\). On BrowseComp-Plus, success improves by \(2.5\) percentage points at \(k=1\) but falls below baseline at \(k=4\), with token usage reaching \(1.43\times\) the baseline. On LiveCodeBench, success similarly declines with replacement depth, from \(68.25\%\) at \(k=1\) to \(57.75\%\) at \(k=4\), while token usage reaches approximately \(1.5\times\) the baseline. We further evaluate different replacement models and find that performance does not improve monotonically with substitute model size (\Cref{sec:model_sensitivity}). Thus, while model replacement can improve performance at shallow depths, its effect depends on the chosen substitute, and deeper replacement may both degrade performance and increase token consumption. 

% \subsection{Prompt Optimization: Performance--Cost Trade-offs and Target Selection}

\noindent\textbf{Prompt optimization.}
Prompt optimization \citep{khattab2024dspy, agrawal2026gepa} modifies agent behavior while preserving models and communication structure. On WorkBench, optimizing the top-\(k\) agents improves success by \(10.5\text{--}15.0\) percentage points for \(k=1\text{--}3\); on PlanCraft, the gain peaks at \(5\) points at \(k=3\); and on LiveCodeBench, optimizing the top two agents raises success from approximately \(64\%\) to \(70\%\), while token usage increases from \(7\)M to \(9\)M (\Cref{fig:cost_reduction_plancraft}). We find that optimizing more agents does not consistently improve performance. On LiveCodeBench, optimizing the top two agents achieves \(69.8\%\) success, comparable to \(69.3\%\) with four agents, while using \(16\%\) fewer tokens (9.2M vs. 11.0M). On PlanCraft, optimizing three agents outperforms optimizing all agents \((44.2\%\) vs. \(41.5\%)\) with \(25\%\) lower token usage (6.0M vs. 8.0M). Upon examining the optimizer prompts, it often replaces short descriptions with task-specific examples, procedures, and verification steps (\Cref{prompt:gepa_updated}). Overall, our results show that targeted prompt optimization can match or outperform system-wide prompt optimization with substantially lower inference cost.

\vspace{-0.2em}
\begin{tcolorbox}[colback=blue!5!white,top=0.25em, bottom=0.25em, colframe=blue!75!black]
\noindent\textbf{Key takeaway:} Attribution-guided interventions offer the strongest performance--cost trade-offs at modest intervention depths. Deeper interventions provide diminishing returns: ablation increasingly sacrifices performance, while replacement and prompt optimization can increase inference cost without consistent performance gains.
\end{tcolorbox}
\vspace{-0.8em}

\subsection{Can removal-based attribution guide prompt optimization?}
To guide agent selection for prompt optimization, \citet{xia2026hivemind} propose CG-OPO, which computes Shapley values with agent ablation to identify bottleneck agents. However, under our framework, this creates an intervention--attribution mismatch: importance under removal need not reflect prompt-improvement potential. We therefore compare CG-OPO with top-\(k\) selection using LOO and Shapley attribution based on the prompt-optimization defined in our framework (\Cref{eq:intervention_game}), evaluating task success and attribution cost.

\begin{figure}[t!]
    \centering
    \includegraphics[width=1.0\linewidth]{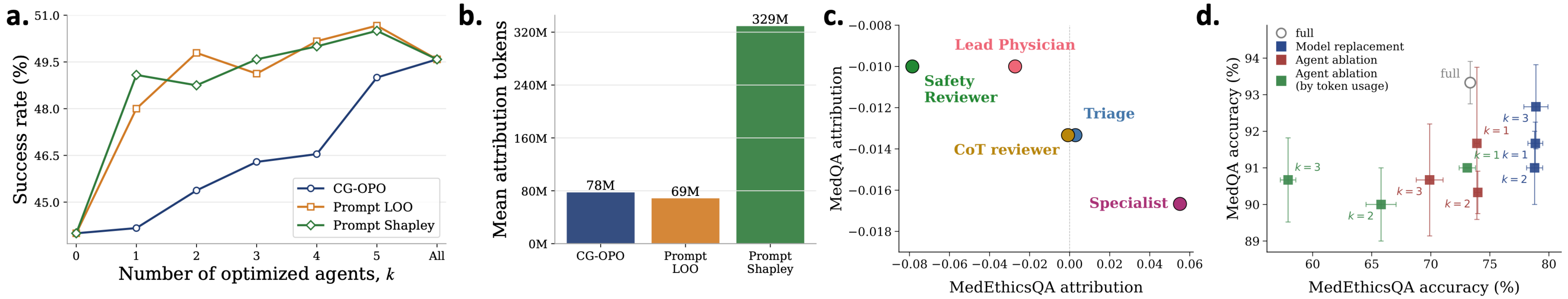}
    \vspace{-1.5em}
    \caption{ \textbf{(a)} Mean task success under top-$k$ prompt optimization using CG-OPO, prompt-LOO, or prompt-Shapley.
    \textbf{(b)} Mean attribution token usage. Panels (a--b) are averaged across datasets and communication topologies.
    \textbf{(c)} Model-replacement attribution for MDTeamGPT on MedQA (\(y\)-axis) and MedEthicsQA (\(x\)-axis).
    \textbf{(d)} MDTeamGPT performance under top-\(k\) model replacement and agent ablation, with agents ranked by the corresponding MedEthicsQA attribution; green denotes a token-usage-based ablation baseline.}
    \label{fig:promptopt_ranking_comparison}
    \vspace{-1em}
\end{figure}

As shown in \Cref{fig:promptopt_ranking_comparison}(a), this mismatch matters in practice. Although ablation-based selection improves mean success, prompt-based attribution performs better at every intervention depth. At \(k=1\), prompt LOO and prompt Shapley achieve \(48.0\%\) and \(49.1\%\) success, respectively, compared with \(44.2\%\) for CG-OPO. The same pattern holds across benchmarks, with the largest gains on WorkBench (\Cref{fig:promptopt_ranking_comparison_gpt54_per_dataset_frac_std}). We also find that prompt LOO tracks prompt Shapley while using \(79\%\) fewer attribution tokens than prompt Shapley and \(12\%\) fewer than CG-OPO (\Cref{fig:promptopt_ranking_comparison}(b)). 

\begin{tcolorbox}[colback=blue!5!white,top=0.25em,
    bottom=0.25em, colframe=blue!75!black]
\noindent \textbf{Key takeaway:} Matching attribution to the intervention improves target selection: prompt LOO identifies more effective optimization targets at lower attribution cost than ablation-guided prompt optimization.
\end{tcolorbox}
\vspace{-0.5em}

\subsection{Evaluation Objectives Shape Agent Attribution: An Audit of MDTeamGPT}

\noindent Prior work typically attributes agents with respect to a single evaluation metric \citep{yang2025s,xia2026hivemind}, leaving open whether agent importance is stable across objectives. Our framework makes this dependence explicit: changing the evaluation metric changes the utility function and therefore induces a distinct attribution game (\Cref{sec:behavior_metrics}). This is particularly important in domains where systems must satisfy multiple behavioral objectives. In healthcare, for example, medical MASs must account for safety and ethical reasoning in addition to diagnostic performance \citep{kim2024mdagents,tang2024medagents,ong2024medical}. To examine whether agent importance changes across such objectives, we audit MDTeamGPT \citep{chen2025mdteamgpt}, a medical consulting MAS using GPT-5-mini as its backbone model, under two evaluation metrics: diagnostic accuracy on MedQA \citep{jin2021disease} and ethical reasoning on MedEthicsQA \citep{wei2025medethicsqa} (\Cref{apx:mdteamgpt_details}).

We compute LOO attribution via model replacement with Qwen 3.5 (27B) under both objectives and find sharply different contribution patterns across metrics. Specifically, we observe a near-complete reversal in agent attribution rankings between diagnostic accuracy and ethics alignment scores (\Cref{fig:promptopt_ranking_comparison}c). For example, replacing the Safety Reviewer decreases MedQA accuracy by \(0.010\) and the ethics score by \(0.08\), indicating that its original backbone benefits both objectives. Conversely, replacing the Specialist modestly decreases accuracy but yields the largest improvement in ethics \((+0.06)\).  These results show that backbone utility in MDTeamGPT is strongly metric-dependent: Agents whose original backbones improve diagnostic performance may reduce ethics alignment.

These attribution differences help identify agents whose original backbones degrade ethical performance while contributing little to diagnostic accuracy: \textbf{Specialist}, \textbf{Triage}, and \textbf{CoT Reviewer}. Replacing these three agents increases the ethics score by $4$ percentage points while preserving MedQA accuracy (\Cref{fig:promptopt_ranking_comparison}d). In contrast, agent ablation does not improve ethics and decreases the ethics score by \(4.6\) percentage points at \(k=3\), with modest reductions in MedQA accuracy.

\begin{tcolorbox}[colback=blue!5!white,top=0.25em,  bottom=0.25em, colframe=blue!75!black]
\noindent \textbf{Key takeaway:} Agent importance depends on the evaluation objective: auditing across behavioral metrics reveals trade-offs and enables interventions that improve desired behaviors.
\end{tcolorbox}
\vspace{-0.5em}

\section{Conclusion}
\vspace{-0.5em}
In this work, we formalize MAS attribution as an intervention-conditioned cooperative game and show that agent importance depends on the intervention protocol, coalition distribution, and target metric. Different interventions therefore answer different attribution questions, while simple LOO often recovers similar rankings to more expensive combinatorial methods. For optimization, modest removal or model replacement often provides the strongest performance--cost trade-off. Overall, attribution is most useful when aligned with the intended intervention and evaluation objective---which agent to change, how to change it, and what system behavior to improve. Together, these results provide a principled basis for analyzing and improving complex multi-agent LLM systems.

% Using this framework, we systematically analyze how attribution distributions, removal protocols, and MAS topologies interact. Empirically, across datasets, we find that removal protocols induce distinct attribution behavior: Agent ablation can emphasize structural bottlenecks, while introspective removal does not reliably approximate agent ablation and remains sensitive to the choice of judge model. We further observe that agent rankings are largely stable across coalition distributions and communication graphs, making leave-one-out an effective and computationally efficient approach. Finally, we propose model replacement as a topology-preserving removal protocol and demonstrate its utility for targeted system interventions, such as identifying roles that can use cheaper models while maintaining downstream task performance. Together, these results provide a principled foundation for evaluating, debugging, and optimizing resource allocation in complex multi-agent LLM workflows.

\clearpage

\section*{AI Use Statement}
In this work, we used generative AI tools to assist with figure generation from experimental results, language and grammar editing, code generation for method implementation, experiment execution, and statistical analysis. All AI-assisted code was reviewed, verified, and tested for correctness by the authors. Separately, LLMs are also an explicit object of study within the proposed framework: the introspective-removal protocol uses an LLM judge to estimate coalition utility, as described in \Cref{sec:removal_protocols}.

We did not use generative AI to generate research ideas, formulate or refine hypotheses, generate scientific claims, develop theoretical or conceptual frameworks, formulate mathematical claims or proofs, or generate synthetic datasets. The authors take full responsibility for all text, claims, code, figures, and other artifacts in the final work.

\section*{Reproducibility Statement}
\Cref{apx:attribution_methods} details the computation of each coalition distribution; \Cref{apx:datasets} describes dataset construction and sources; \Cref{apx:agent_configs} specifies the agent configurations and system implementations; and \Cref{apx:mdteamgpt_details} provides implementation details for MDTeamGPT.

% \section*{Ethical Statement}
% \input{sections/ethical_statement}

% \section*{Reproducibility statement}

\bibliography{iclr2027_conference}
\bibliographystyle{iclr2027_conference}

\clearpage
\appendix
\section*{Appendix}
\section{Attribution Methods and Coalition Weighting}\label{apx:attribution_methods}
We provide formal definitions of the attribution methods evaluated in our experiments.

% \paragraph{Leave-One-Out (LOO)}
% LOO attribution measures the marginal contribution of an agent by comparing the utility of the full coalition with the utility of the coalition from which the agent is removed. Formally, the LOO score for agent $a_i \in \mathcal{A}$ is defined as:
% \begin{equation}\label{eq:loo}
% \phi_{\mathrm{LOO}}(\nu, \mathcal{A})_i = \nu(\mathcal{A}) - \nu(\mathcal{A} \setminus \{a_i\}),
% \end{equation}

\paragraph{Leave-One-Out (LOO)}
LOO evaluates each agent by intervening on that agent while leaving all others unchanged. Under our formulation, this corresponds to placing all attribution weight on the empty intervention set, $S=\varnothing$. The LOO attribution for agent $a_i$ under intervention $\mathbf{b}$ is
\begin{equation}\label{eq:loo}
\phi_{\mathrm{LOO},i}^{\mathbf{b}} = \nu_{\mathbf{b}}(\{a_i\}) - \nu_{\mathbf{b}}(\varnothing).
\end{equation}
Thus, LOO measures the marginal effect of intervening on $a_i$ relative to the original system, with all other agents unchanged. Positive values indicate that the intervention improves utility, whereas negative values indicate that it degrades utility.

% \paragraph{Banzhaf Value \citep{dubey1979mathematical}} 
% The Banzhaf value is a \textit{semivalue} that measures an agent’s average marginal contribution over all possible coalitions that exclude it, assigning equal weight to each. It satisfies all Shapley axioms except \textit{efficiency}. Formally:
% \begin{equation}\label{eq:banzhaf}
% \phi_{\mathrm{Banzhaf}}(\nu, \mathcal{A})_i = \mathbb{E}_{S \sim w_{\mathrm{Banzhaf}}} \left[\nu(S \cup \{a_i\}) - \nu(S)\right],
% \end{equation}
% where \( w_{\mathrm{Banzhaf}} \) denotes the uniform distribution over subsets \( S \subseteq \mathcal{A} \setminus \{a_i\} \).

\paragraph{Shapley Value \citep{shapley1953value}}
The Shapley value averages the effect of intervening on an agent across all intervention coalitions, weighting each intervention set according to its size. For an agent $a_i \in \mathcal{A}$ with $n = |\mathcal{A}|$, the Shapley attribution under intervention $\mathbf{b}$ is
\begin{equation}
\label{eq:shapley}
\phi^{\mathbf{b}}_{\mathrm{Shapley},i}
=
\frac{1}{n}
\sum_{S \subseteq \mathcal{A}\setminus\{a_i\}}
\binom{n-1}{|S|}^{-1}
\left[
\nu_{\mathbf{b}}(S\cup\{a_i\})
-
\nu_{\mathbf{b}}(S)
\right].
\end{equation}
The weighting assigns equal total mass to each intervention-set size. Thus, unlike LOO, which evaluates the intervention only when all other agents remain unchanged, Shapley averages its effect across intervention contexts of all sizes.

\paragraph{Owen Value \citep{owen1977essays}}
The Owen value extends the Shapley value to settings with a predefined coalition structure. Let $\mathcal{G}=\{G_1,\ldots,G_m\}$ be a partition of agents into groups, and let $G(i)$ denote the group containing agent $a_i$. The Owen value averages the effect of intervening on $a_i$ over both the order in which groups join the coalition and the order in which agents within $G(i)$ join. Let $U_T=\bigcup_{G\in T}G$ denote the agents contained in the preceding groups. The Owen attribution under intervention $\mathbf{b}$ is
\begin{equation}
\label{eq:owen}
\phi^{\mathbf{b}}_{\mathrm{Owen},i}
=
\sum_{T \subseteq \mathcal{G}\setminus\{G(i)\}}
\sum_{S \subseteq G(i)\setminus\{a_i\}}
w(T,S)
\left[
\nu_{\mathbf{b}}(U_T \cup S \cup \{a_i\})
-
\nu_{\mathbf{b}}(U_T \cup S)
\right],
\end{equation}
where
\begin{equation}
w(T,S)
=
\frac{|T|!(m-|T|-1)!}{m!}
\frac{|S|!(|G(i)|-|S|-1)!}{|G(i)|!}.
\end{equation}
Thus, Owen attribution measures the effect of intervening on an agent while respecting a predefined grouping structure over agents.

\paragraph{Myerson Value \citep{myerson1977graphs}}
The Myerson value extends the Shapley value to games with graph-structured
communication constraints. Let $g=(\mathcal{A},E_g)$ be an undirected graph over agents. We derive $g$ from the MAS communication graph $\Gamma$ by ignoring edge direction, connecting two agents whenever a communication link exists in either direction. This yields a star graph for the centralized topology and complete graphs for the decentralized and hybrid topologies. For the independent topology, which has no inter-agent communication, we use a complete graph so that Myerson reduces to Shapley.

For any coalition $S\subseteq\mathcal{A}$, let $S/g$ denote the connected
components induced by $S$ in $g$. Because $\nu_{\mathbf b}(\varnothing)$ need not be zero, we subtract this baseline before summing component utilities to avoid counting it once per connected component. We define the graph-restricted intervention-conditioned utility as
\begin{equation}
\label{eq:myerson_restricted_game}
\nu_{\mathbf b,g}(S)
=
\sum_{C\in S/g}
\left[
\nu_{\mathbf b}(C)
-
\nu_{\mathbf b}(\varnothing)
\right].
\end{equation}
This reduces to the standard graph restriction when $\nu_{\mathbf b}(\varnothing)=0$. The Myerson attribution for agent $a_i$ under intervention $\mathbf b$ is then the Shapley attribution of this graph-restricted game:
\begin{equation}
\label{eq:myerson}
\phi^{\mathbf b}_{\mathrm{Myerson},i}
=
\frac{1}{n}
\sum_{S\subseteq\mathcal{A}\setminus\{a_i\}}
\binom{n-1}{|S|}^{-1}
\left[
\nu_{\mathbf b,g}(S\cup\{a_i\})
-
\nu_{\mathbf b,g}(S)
\right].
\end{equation}

\section{Datasets}\label{apx:datasets}
\subsection{PlanCraft}
PlanCraft \footnote{\url{https://github.com/gautierdag/plancraft}} is a Minecraft-inspired planning benchmark for evaluating LLM agents on structured crafting tasks \citep{dagan2024plancraft}. It provides text-only and multimodal interfaces based on the Minecraft crafting GUI, requiring agents to reason over inventories, recipes, intermediate dependencies, and actions. The benchmark contains 1,145 training, 570 validation, and 580 test examples, with tasks grouped into five difficulty levels and an additional set of intentionally unsolvable tasks. It also includes the Minecraft Wiki as an external knowledge source for tool use and retrieval-augmented reasoning. In our pipeline, we use its first 400 instances in the fixed dataset order: 100 for attribution estimation, 100 for prompt optimization, and a disjoint set of 200 for evaluating interventions. The environment exposes four tools, \texttt{search}, \texttt{move}, \texttt{smelt}, and \texttt{impossible}. Every agent can \texttt{search} recipes, while the action tools are assigned by role: the mover has \texttt{move}, the smelter \texttt{smelt}, the crafter both, and only the validator may additionally declare a target \texttt{impossible}.

\subsection{WorkBench}

WorkBench\footnote{\url{https://github.com/olly-styles/WorkBench}}  is a workplace-task benchmark for evaluating agents in realistic business environments \citep{styles2024workbench}. It provides a sandbox with five databases and 26 tools covering analytics, calendar management, CRM, email, and project management. The benchmark contains 690 tasks from 69 human-curated templates, organized into six categories: Analytics, Calendar, CRM, Email, Project Management, and Multi-domain. WorkBench uses outcome-centric evaluation, where success is determined by the final database state. We draw a stratified sample of 400 tasks proportional to the domain composition of the benchmark and force-include the ten tasks that span all five domains: 100 tasks for attribution estimation
(70 single-domain and 30 cross-domain), 100 for prompt optimization, and a disjoint set of 200 for evaluating interventions (134 single-domain and 66 cross-domain). Agents have access to all 26 native WorkBench tools and a \texttt{submit} action.

\subsection{BrowseComp-Plus}
BrowseComp-Plus\footnote{\url{https://github.com/texttron/BrowseComp-Plus}} is a deep-research benchmark for evaluating search-augmented agents under controlled retrieval conditions \citep{chen2025browsecomp}. It replaces live web search with a fixed corpus, enabling reproducible evaluation of retrieval, evidence use, reasoning, and answer synthesis. The benchmark contains 830 queries and 100,195 documents, with human-verified supporting documents and challenging negatives for each query. In our pipeline,  we draw a random subset of 400 queries with fixed seeds: 100 for attribution estimation, 100 for prompt optimization, and a disjoint set of 200 for evaluating interventions, while retaining the full document corpus. Agents
search the corpus with a dense retriever (Qwen3-Embedding-4B) that returns the top-5 documents with snippets per query, can retrieve any document in full by ID, and submit their final answer with \texttt{done}. Agents use the \texttt{search\_documents} and \texttt{retrieve\_document} tools, signal completion with \texttt{done}, and are graded by \texttt{gpt-4o-mini} following the BrowseComp grading protocol.

\subsection{LiveCodeBench}
LiveCodeBench\footnote{\url{https://github.com/LiveCodeBench/LiveCodeBench}} is a continuously updated benchmark for evaluating code generation on competitive programming problems \citep{jain2025livecodebench}. It collects problems from LeetCode, AtCoder, and Codeforces and evaluates generated solutions against hidden test cases. We use 400 problems from the \texttt{code\_generation\_lite} releases v4--v6 (443 problems in total, contests from 2023-08 to 2025-04), sampled with fixed seeds: 200 problems from the v5 and v6 releases (100 for attribution estimation and 100 for prompt optimization) and a disjoint set of 200 from the remaining problems of v4--v6 for evaluating interventions. Agents share a single solution buffer and may run the problem's public example tests; a solution is graded on the hidden test cases with a 6-second timeout per test, and success is measured by pass@1.

\clearpage
\section{Agent Configuration}\label{apx:agent_configs}

We evaluate each specialist partition under four multi-agent architectures that vary in their communication structure. Across architectures, specialization is implemented with the same two mechanisms: agent-specific personas inserted into each worker's system prompt, and agent-specific tool whitelists applied during environment initialization. 

\subsection{Architectures.}
Each architecture uses five base workers for PlanCraft and WorkBench, and four base workers for BrowseComp-Plus and LiveCodeBench.

\paragraph{Independent.}
Workers run in parallel without inter-agent communication or an orchestrator. Each worker attempts the full task independently with at least three and at most five iterations. The final answer returns the first successful attempt, falling back to the first completed attempt.

\paragraph{Centralized.} A single lead agent decomposes the task, assigns one subtask to each worker slot, sends per-round coordination messages, and synthesizes the final answer once its stopping criterion is met. Workers observe only their own conversation and the orchestrator's messages; there is no direct peer-to-peer channel. The system runs for at most five coordination rounds, with three worker iterations per round.

\paragraph{Decentralized.} Workers act as peers without an orchestrator. They exchange findings after each round and terminate when a 0.7 consensus threshold is reached. The system runs for at most five rounds, with three worker iterations per round.

\paragraph{Hybrid.}  This architecture combines centralized orchestration with direct peer-to-peer communication among workers. Workers receive both orchestrator messages and teammates' findings each round. The round and timeout budgets match the centralized setting.

\subsection{Orchestrator Agent}
The centralized and hybrid architectures instantiate a single lead agent alongside the worker slots. Unlike workers, the lead agent has no access to environment action tools; it does not call tools such as \texttt{move}, \path{search_documents}, or \path{calendar_create_event}. Its interface is plain LLM generation, used for planning, coordination, stopping, and synthesis.

At the start of each task, the lead agent produces a JSON \texttt{OrchestrationPlan} containing one \{\texttt{agent\_id}, \texttt{objective}, \texttt{focus}\} subtask per worker slot. If parsing or validation fails, the system falls back to a generic per-slot plan. During execution, the lead agent reads each worker's accumulated findings and the team-wide context, then sends short coordination messages before the next worker iteration. After each round, it decides whether the team has sufficient evidence to answer or should continue, subject to the global round and timeout budgets. Once stopping is triggered or the budget is exhausted, the lead agent synthesizes the final answer from the full findings buffer.

The lead agent has full read access to worker findings. In the centralized architecture, workers receive only the lead agent's coordination messages and have no direct peer channel. In the hybrid architecture, the same lead agent is used, but workers also receive peer findings between rounds. The independent and decentralized architectures do not instantiate a lead agent.

\subsection{Worker Agent: PlanCraft}\label{sec:agent-cfg-plancraft}

PlanCraft exposes four action tools: \texttt{search}, \texttt{move}, \texttt{smelt}, and \texttt{impossible}. Because the action space is small and crafting actions are largely sequential, all worker slots retain the universally useful \texttt{search} tool, while the remaining action tools are distributed with deliberate overlap.

\noindent\textbf{Agent 1: Recipe Scout.} Tools: \texttt{search}. Retrieves recipes, ingredients, and intermediate steps, but cannot modify the inventory.

\noindent\textbf{Agent 2: Inventory Mover.} Tools: \texttt{search}, \texttt{move}. Places items in the crafting grid and can complete recipes requiring only inventory movement, but cannot smelt.

\noindent\textbf{Agent 3: Smelter.} Tools: \texttt{search}, \texttt{smelt}. Converts raw materials using the furnace, but cannot move items within the inventory.

\noindent\textbf{Agent 4: Crafter.} Tools: \texttt{search}, \texttt{move}, \texttt{smelt}. Has the full constructive action set and is responsible for completing feasible recipes.

\noindent\textbf{Agent 5: Validator.} Tools: \texttt{search}, \texttt{move}, \texttt{smelt}, \texttt{impossible}. Has the same constructive tools as the Crafter and is the only slot allowed to declare a task impossible.

\subsection{Worker Agent: WorkBench}\label{sec:agent-cfg-workbench}

WorkBench provides 26 native tools across five domains: calendar, email, analytics, project management, and CRM. It also includes shared company-directory lookup tools and a \texttt{submit} terminator. Our specialist partition assigns three single-domain specialists and two CRM--email generalists, encouraging cross-domain coordination on multi-domain tasks. Tool whitelists are specified as \texttt{fnmatch} globs over registered tool names. Every worker has access to the company-directory tools and \texttt{submit}.

\noindent\textbf{Agent 1: Calendar Coordinator.} Tools: \path{calendar_*}, \path{company_directory_*}, \texttt{submit}. Handles scheduling, meetings, and availability constraints.

\noindent\textbf{Agent 2: Project Manager.} Tools: \path{project_management_*}, \path{company_directory_*}, \texttt{submit}. Creates, updates, searches, and closes project-management tasks.

\noindent\textbf{Agent 3: Analytics Analyst.} Tools: \path{analytics_*}, \path{company_directory_*}, \texttt{submit}. Handles visitor counts, traffic sources, session metrics, and plot generation.

\noindent\textbf{Agents 4--5: CRM--Email Generalists.} Tools: \path{crm_*}, \path{email_*}, \path{company_directory_*}, \texttt{submit}. These two identically scoped agents share the customer-lookup and email-communication interface. In centralized and hybrid settings, the orchestrator coordinates them to reduce duplicated actions; in independent and decentralized settings, they act as a redundant pair.

For centralized and hybrid WorkBench runs, the orchestrator must use the literal slot identifiers \texttt{agent\_1}, \ldots, \texttt{agent\_5} in the subtask \texttt{agent\_id} field. Semantic names such as ``calendar'' or ``analyst'' are not accepted, so role information is encoded only in the \texttt{focus} field.

\subsection{Worker Agent: BrowseComp-Plus}\label{sec:agent-cfg-browsecomp}

BrowseComp-Plus exposes three tools: \texttt{search\_documents}, \texttt{retrieve\_document}, and \texttt{done}. Since partitioning this small tool set would leave some workers without essential actions or simply duplicate the same allocation, all four workers share the full tool set. Specialization is instead encoded through personas that decompose BrowseComp-style queries into complementary verification axes.

\noindent\textbf{Agent 1: Candidate Enumerator.} Tools: \path{search_documents}, \path{retrieve_document}, \texttt{done}. Generates a ranked shortlist of candidate entities using concrete searchable hooks such as years, awards, roles, or events. Prioritizes recall and reports document-backed candidates for verification.

\noindent\textbf{Agent 2: Biographical Verifier.} Tools: \path{search_documents}, \path{retrieve_document}, \texttt{done}. Confirms or rejects biographical facts for each candidate, including dates, locations, family background, education, and other identity constraints.

\noindent\textbf{Agent 3: Relationship and Event Tracker.} Tools: \path{search_documents}, \path{retrieve_document}, \texttt{done}. Verifies family relationships and time-bound events, such as spouses, children, dated incidents, and constraints tied to the target entity.

\noindent\textbf{Agent 4: Answer Synthesizer.} Tools: \path{search_documents}, \path{retrieve_document}, \texttt{done}. Integrates teammates' findings, issues targeted and disambiguating queries, and commits the final answer once all question constraints are supported.

The \texttt{done} terminator is available to all BrowseComp-Plus workers because any worker may identify a sufficiently supported answer. In centralized and hybrid settings, the orchestrator selects the committed finding used as the final response; in independent and decentralized settings, the registered answer is determined by aggregation or consensus over worker outputs. The full agent prompt templates are provided in \Cref{apx:prompt_template}.

\subsection{Worker Agent: LiveCodeBench}\label{sec:agent-cfg-lcb}

LiveCodeBench provides four native tools over a single candidate solution: \texttt{write\_solution} replaces the candidate program, \texttt{read\_solution} returns it, \texttt{run\_public\_tests} executes it against the problem's public examples, and \texttt{submit} declares it final. Following \cite{hong2023metagpt}, our specialist partition assigns one role per stage of a plan--code--test--review workflow, with role-exclusive tools: only one agent can edit the solution and only one can execute it, so progress requires information to pass between roles. Tool whitelists are specified as exact tool names. Every worker has access to \texttt{read\_solution} and \texttt{submit}.

\noindent\textbf{Agent 1: Planner.} Tools: \texttt{read\_solution}, \texttt{submit}. Analyzes the problem, identifies the algorithm, complexity bounds, edge cases, and I/O format, provides the Coder with a plan, and reviews the current solution against that plan.

\noindent\textbf{Agent 2: Coder.} Tools: \texttt{write\_solution}, \texttt{read\_solution}, \texttt{submit}. Turns the plan into complete Python 3 code, preserves starter-code signatures, and fixes failures reported by the Tester. This is the only agent that can modify the solution.

\noindent\textbf{Agent 3: Tester.} Tools: \texttt{run\_public\_tests}, \texttt{read\_solution}, \texttt{submit}. Runs the public examples against the current solution, diagnoses failures such as wrong outputs, exceptions, and timeouts, and reports the required changes. This is the only agent that can execute code, and it cannot edit it.

\noindent\textbf{Agent 4: Reviewer.} Tools: \texttt{read\_solution}, \texttt{submit}. Inspects the current solution for logic errors, off-by-one mistakes, unhandled edge cases, and performance problems on large inputs, and reports concrete fixes. It can neither edit nor execute code.

In decentralized, centralized, and hybrid settings, all agents operate on one shared solution buffer per task; in the independent setting, each agent maintains its own buffer. A \texttt{submit} call is an agent's vote that the current candidate is final and ends only that agent's turn for the current round, so the orchestrator may dispatch the same agent again in later rounds. The final candidate is graded against the hidden test cases.

For centralized and hybrid LiveCodeBench runs, the orchestrator must use the literal slot identifiers \texttt{agent\_1}, \ldots, \texttt{agent\_4} in the subtask \texttt{agent\_id} field. Semantic names such as ``planner'' or ``coder'' are not accepted, so role information is encoded only in the \texttt{focus} field.

\subsection{Implementation Details}

We implement our experiments using the agent-scaling framework\footnote{\url{https://github.com/ybkim95/agent-scaling}} in Python 3.11, with LiteLLM\footnote{\url{https://github.com/BerriAI/litellm}} version 1.73.6 and LangChain version 0.3.26 \footnote{\url{https://github.com/langchain-ai/langchain}}. All tasks are executed in a standard Python 3.11 environment. The BrowseComp-Plus, PlanCraft, WorkBench, and LiveCodeBench datasets are run directly, without sandboxing. We summarize the experimental configuration below.

\paragraph{Model Configuration.}
All experiments use the backbone models described in \Cref{sec:mas_configs}. LLM requests are routed through the OpenRouter \footnote{\url{https://openrouter.ai/}} API using LiteLLM. We use a temperature of 0.0, corresponding to greedy decoding for pass@1 evaluation. We do not override top-p sampling, and we use the provider's default server-side maximum token limit.

\paragraph{Framework Hyperparameters.}
We retain the default hyperparameters of the framework. The outer coordination loop is limited to 5 rounds. Each worker agent is assigned at most 5 action iterations, with the minimum number of iterations set to 3. The memory of collected findings is capped at 100 entries. Each task has a wall-clock time limit of 600 seconds, and each worker has an individual timeout of 300 seconds.

\paragraph{Cost Tracking.}
We record token usage and monetary cost using LiteLLM's built-in callbacks. These callbacks capture the exact prompt-token and completion-token counts returned by the OpenRouter API for every model call.

\section{Additional Details for MDTeamGPT}\label{apx:mdteamgpt_details}

MDTeamGPT \citep{chen2025mdteamgpt} is configured as a multi-agent clinical consultation framework consisting of specialist doctor agents and auxiliary reviewer agents. The system follows a multi-round consultation workflow based on residual discussion and consensus aggregation. In this workflow, a Primary Care Doctor first triages the patient case and assigns it to appropriate specialist agents. The specialists then discuss the case over multiple rounds, incorporating prior-round information through a Historical Shared Pool. A Lead Physician summarizes and organizes the discussion, while reviewer agents verify safety and archive consultation experience for future use.

\vspace{1em}
\noindent\textbf{Primary Care Doctor.}
The Primary Care Doctor serves as the triage agent. Given the patient's symptoms, signs, and medical history, this agent assigns the case to appropriate specialist doctors for further consultation.

\vspace{1em}
\noindent\textbf{Specialist Doctors.}
Specialist doctors include roles such as General Internal Medicine Doctor, Pediatrician, and Radiologist. Each specialist provides diagnoses and treatment recommendations based on the patient case. During multi-round discussions, specialists refer to the Historical Shared Pool to incorporate insights from previous rounds before selecting an optimal treatment plan.

\vspace{1em}
\noindent\textbf{Lead Physician.}
The Lead Physician organizes and summarizes the diagnoses and treatment recommendations generated by the specialist agents. It categorizes specialist responses into four relation types: Consistency, Conflict, Independence, and Integration. These structured summaries are stored in the Historical Shared Pool, helping maintain an organized discussion process and reducing cognitive load across rounds.

\vspace{1em}
\noindent\textbf{Chain-of-Thought Reviewer.}
The Chain-of-Thought Reviewer extracts structured reasoning traces from each doctor's diagnostic process. Depending on whether the consultation outcome is correct or incorrect, the extracted statements are archived into either the Correct Answer Knowledge Base, denoted as CorrectKB, or the Chain-of-Thought Knowledge Base, denoted as ChainKB.

\vspace{1em}
\noindent\textbf{Safety and Ethics Reviewer.}
The Safety and Ethics Reviewer verifies and refines the final consultation outcome. It filters harmful, unsafe, or unethical suggestions to ensure that the final diagnosis and treatment recommendation satisfy safety and ethical constraints.

\subsection{Datasets: MedQA and MedEthicsQA}\label{sec:dataset-cfg-mdteamgpt}

We evaluate MDTeamGPT on two medical reasoning benchmarks: MedQA\footnote{\url{https://github.com/jind11/MedQA}}\citep{jin2021disease}, using 100 sampled USMLE-style multiple-choice questions with four or five answer options, and MedEthicsQA\footnote{\url{https://github.com/JianhuiWei7/MedEthicsQA}}\citep{wei2025medethicsqa}, using 100 multiple-choice questions and 100 open-ended questions. MedQA assesses medical knowledge and practical clinical reasoning, while MedEthicsQA evaluates the system's ability to reason about medical ethics in both answer-selection and free-form response settings. For multiple-choice questions in both datasets, we report accuracy. For MedEthicsQA open-ended questions, we evaluate the final response using an LLM-as-a-judge protocol that compares the generated answer against the reference answer and assigns a binary correctness label; we report the resulting judge-based accuracy using the same formula.

\subsection{Implementation Details}\label{sec:implementation-mdteamgpt}
Our implementation settings follow \cite{chen2025mdteamgpt}\footnote{\url{https://github.com/KaiChenNJ/MDTeamGPT}} and are summarized below.

\paragraph{Experimental Environment.}
We implement the system in Python 3.13 using LangGraph for multi-agent orchestration. We implement vector similarity search for knowledge base retrieval with FAISS.

\paragraph{Backbone Model Configuration.}
In our experiments, MDTeamGPT uses GPT-5-mini as the backbone model for all agent configurations. For embedding textual content into the vector spaces of the knowledge bases, we use OpenAI's \texttt{text-embedding-3-small} model.

\paragraph{Framework Hyperparameters.}
The system conducts multi-round specialist discussions until all specialist doctor agents reach consensus, subject to a maximum of 6 rounds. If consensus is not reached within this limit, the final decision is determined by majority vote. For retrieval-based prompt enhancement from the knowledge bases, the system selects the top $K=5$ most similar cases.

\paragraph{Temperature Settings.}
Specialist doctor agents use a temperature of 0.7 to encourage diverse clinical perspectives. The Lead Physician, Safety and Ethics Reviewer, and CoT Reviewer all use a temperature of 0.0 for deterministic outputs.

\paragraph{Knowledge Base Configuration.}
MDTeamGPT maintains two experience repositories: the Correct Answer Knowledge Base, CorrectKB, and the Chain-of-Thought Knowledge Base, ChainKB. In the main experiments, the framework stores consultation experience from 600 consultation rounds.

\clearpage

\section{Additional Results}
\begin{table}[t!]
    \centering
    \caption{Agreement ($R^{2}$ / Spearman $\rho$) between introspective and agent-ablation coalition values across four benchmarks, averaged over 3 runs. Best results are shown in bold. $\mathrm{ns}$ denotes $p \ge 0.05$ for Spearman $\rho$.}
    \resizebox{0.65\columnwidth}{!}{%
    \begin{tabular}{lcccc}
    \toprule
    \textbf{Judge} & \textbf{Independent} & \textbf{Decentralized} & \textbf{Centralized} & \textbf{Hybrid} \\
    \midrule
    \multicolumn{5}{l}{\textbf{PlanCraft}} \\
    \midrule
    DeepSeek-V4-Flash  & $-1.05$ / 0.47 & $-2.03$ / 0.52 & $-11.93$ / 0.80 & $-0.99$ / 0.14 \\
    DeepSeek-V4-Pro    & $-0.77$ / 0.43 & $-1.12$ / 0.56 & $-21.46$ / 0.69 & $-2.16$ / 0.01 \\
    GPT-5-mini         & 0.38 / 0.47 & 0.78 / 0.72 & $-15.61$ / 0.74 & $-2.16$ / $-0.08$ \\
    GPT-5.4            & $-17.90$ / 0.18 & $-10.21$ / 0.56 & $-39.48$ / 0.67 & $-7.79$ / $-0.06$ \\
    \midrule
    \multicolumn{5}{l}{\textbf{WorkBench}} \\
    \midrule
    DeepSeek-V4-Flash  & $-5.84$ / 0.92 & $-10.07$ / 0.59 & $-5.26$ / 0.85 & $-3.58$ / 0.63 \\
    DeepSeek-V4-Pro    & $-4.41$ / 0.92 & $-6.16$ / 0.64 & $-2.20$ / 0.71 & $-1.71$ / 0.59 \\
    GPT-5-mini         & $-2.54$ / 0.93 & $-1.10$ / 0.84 & $-0.38$ / 0.81 & $-0.32$ / 0.61 \\
    GPT-5.4            & $-3.24$ / 0.93 & $-11.30$ / 0.63 & $-12.09$ / 0.76 & $-11.21$ / 0.66 \\
    \midrule
    \multicolumn{5}{l}{\textbf{BrowseComp-Plus}} \\
    \midrule
    DeepSeek-V4-Flash  & $-1532$ / $-0.73$ & 0.27 / 0.71 & 0.08 / 0.92 & $-4.81$ / 0.18$^{\mathrm{ns}}$ \\
    DeepSeek-V4-Pro    & $-1552$ / $-0.72$ & 0.31 / 0.75 & 0.69 / 0.81 & $-2.58$ / 0.23$^{\mathrm{ns}}$ \\
    GPT-5-mini         & $-2169$ / $-0.74$ & 0.22 / 0.76 & $-1.20$ / 0.36$^{\mathrm{ns}}$ & $-1.04$ / $-0.27$$^{\mathrm{ns}}$ \\
    GPT-5.4            & $-2551$ / $-0.77$ & $-0.42$ / 0.72 & $-0.82$ / 0.78 & $-0.14$ / 0.13$^{\mathrm{ns}}$ \\
    \midrule
    \multicolumn{5}{l}{\textbf{LiveCodeBench}} \\
    \midrule
    DeepSeek-V4-Flash  & 0.89 / 0.87 & $-0.02$ / 0.83 & $-0.50$ / 0.71 & $-0.08$ / 0.62 \\
    DeepSeek-V4-Pro    & 0.96 / 0.80 & 0.50 / 0.82 & $-0.39$ / 0.82 & 0.25 / 0.83 \\
    GPT-5-mini         & 0.96 / 0.76 & 0.30 / 0.77 & $-0.79$ / 0.84 & $-0.26$ / 0.80 \\
    GPT-5.4            & 0.79 / 0.82 & $-0.31$ / 0.77 & $-2.26$ / 0.64 & $-1.36$ / 0.54 \\
    \bottomrule
    \end{tabular}
    }
    \label{tab:coalition-agreement-combined}
\end{table}

\subsection{Fidelity of Introspective Removal}

\begin{figure}[h!]
    \centering
    \includegraphics[width=0.8\linewidth]{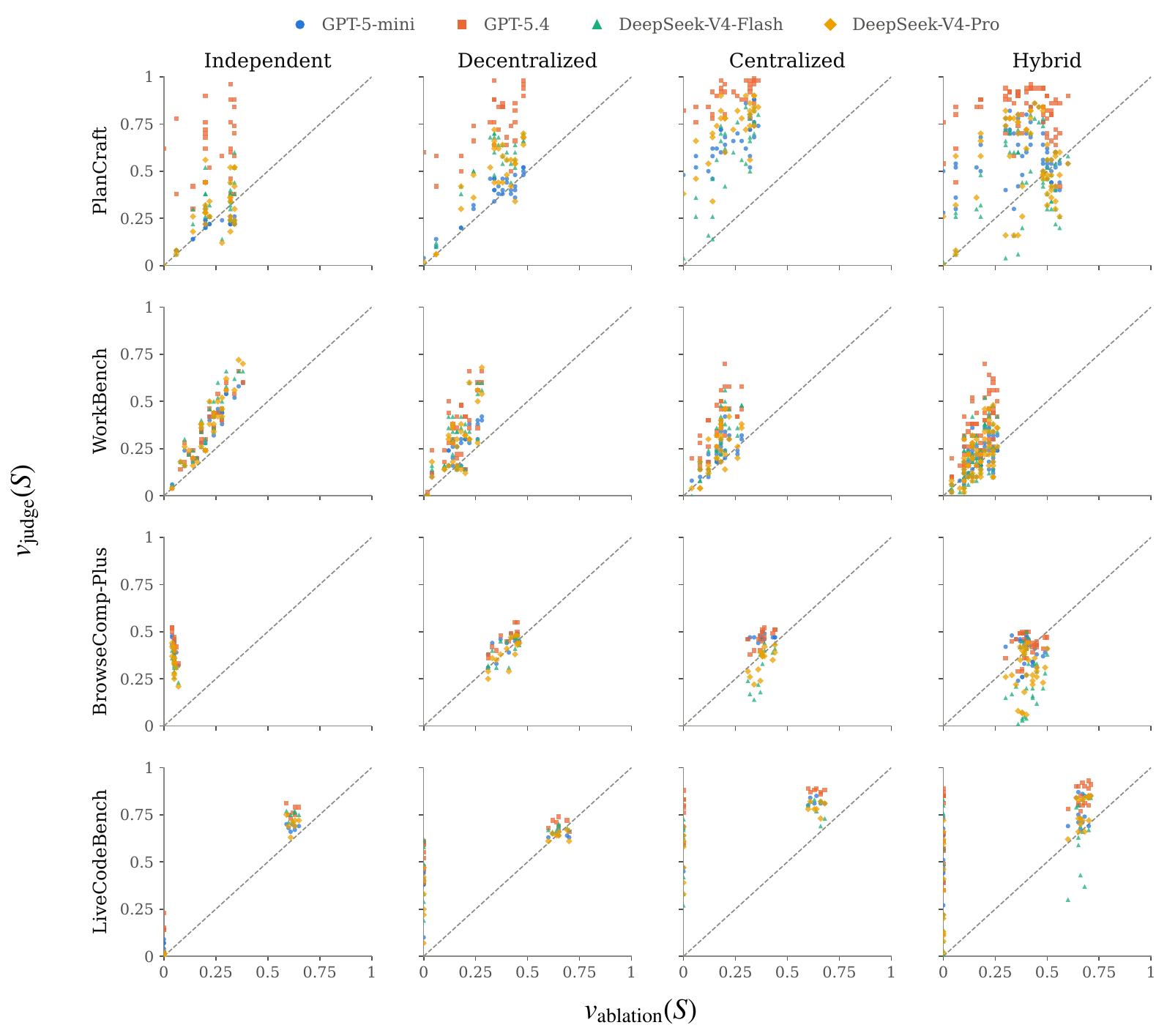}
    \caption{{Judge-estimated versus actual coalition values across datasets and collaboration patterns.} Each point denotes a coalition, with agent-ablation utility on the \(x\)-axis and introspective-removal utility on the \(y\)-axis; the dashed line indicates perfect agreement.}
\label{fig:judge_vS_scatter_4bench}
\end{figure}

\Cref{fig:judge_vS_scatter_4bench} compares judge-estimated and agent-ablation coalition utilities, while \Cref{tab:coalition-agreement-combined} reports the corresponding $R^2$ and Spearman $\rho$. Across benchmarks, judges often preserve the coarse ordering of coalitions while systematically shifting or dispersing their estimated utilities. This explains why moderate or high $\rho$ can coexist with strongly negative $R^2$. Fidelity also depends on the spread of true coalition utilities: agreement is weakest when utilities are tightly concentrated and strongest when coalitions form well-separated groups.

\noindent\textbf{PlanCraft.}
Rank agreement varies substantially by topology and judge. Decentralized and centralized systems retain moderate-to-high $\rho$, while centralized and hybrid systems have uniformly negative $R^2$, consistent with the large upward shifts in \Cref{fig:judge_vS_scatter_4bench}. Thus, coalition ordering can remain partially preserved even when absolute utilities are poorly calibrated.

\noindent\textbf{WorkBench.}
Rank agreement is strongest in the independent topology and remains moderate elsewhere ($\rho=0.59$--$0.85$), yet all $R^2$ values are negative. The dominant error is therefore a systematic shift in coalition utility rather than a complete loss of rank information.

\noindent\textbf{BrowseComp-Plus.}
Agreement varies strongly by topology. In independent systems, ablation utilities are uniformly low while judges assign higher utility \(0.2\)--\(0.5\), yielding negative rank correlation for all judges (\(\rho=-0.77\) to \(-0.72\)). Agreement improves with communication: decentralized systems reach \(\rho=0.71\)--\(0.76\) with the best absolute agreement (\(R^2=-0.7\) to \(0.3\)), and centralized systems reach \(\rho=0.78\)--\(0.92\) for all but GPT-5-mini, despite negative \(R^2\). Hybrid systems remain weak (\(\rho=-0.27\) to \(0.23\)).

\noindent\textbf{LiveCodeBench.}
Half of the coalitions have zero ablation utility because coalitions without the Coder cannot solve the task. Independent judges largely recover this structure: Coder-less coalitions receive utilities of $0.00$--$0.10$, while Coder-containing coalitions lie at $0.6$--$0.8$, yielding the strongest absolute agreement ($R^2=0.79$--$0.96$). Under communication, this separation weakens as Coder-less coalitions receive substantially higher estimates. Rank agreement therefore remains relatively strong ($\rho=0.54$--$0.84$), while $R^2$ becomes negative because many zero-utility coalitions are overestimated.

Overall, judge capability does not monotonically improve either absolute or rank agreement. Stronger judges therefore do not reliably reproduce the coalition utilities obtained by actual ablation.

\subsubsection{Effectiveness of Judge Variants}
\begin{table}[t]
\centering
\caption{
Sensitivity of introspective removal to judge design (GPT-5-mini) on PlanCraft.  Each cell reports agreement ($R^{2}$ / Spearman $\rho$) against agent-ablation coalition values, averaged over three runs. Best results within each topology are bolded.
}
\small
\setlength{\tabcolsep}{5pt}
\resizebox{0.78\columnwidth}{!}{%

\begin{tabular}{lcccc}
\toprule
Judge variant
& \textbf{Indep.}
& \textbf{Decent.}
& \textbf{Cen.}
& \textbf{Hybrid} \\
\midrule
Baseline & 0.38 / 0.47 & 0.78 / 0.72 & $-15.61$ / 0.74 & $-2.16$ / $-0.08$ \\
\midrule
G-Eval w counterfactual rubrics \citep{liu2023g}
& $-1.48$ / 0.55
& $-2.28$ / 0.46
& $-52.03$ / 0.61
& $-33.03$ / 0.63 \\

Self-refinement \citep{madaan2023self}
& $-0.57$ / 0.54
& $-1.15$ / 0.46
& $-43.58$ / 0.61
& $-29.19$ / 0.63 \\

Self-consistency \cite{wang2022self}
& 0.38 / {0.66}
& {0.34} / {0.57}
& $-22.17$ / 0.77
& $-17.12$ / {0.65} \\
\bottomrule
\end{tabular}
}
\label{tab:judge_design_sensitivity}
\end{table}

In this section, we select the strongest introspective judge (\Cref{sec:removal_implementation})(GPT-5-mini) as the backbone and implement three variants: \emph{Rubric-based Chain-of-Thought (CoT) prompting}\citep{liu2023g}, augmented with an explicit counterfactual rubric; \emph{Self-Consistency}\citep{wang2022self}, which majority-votes \(K{=}3\) perturbed baseline judgments; and \emph{Self-Refinement}\citep{madaan2023self}, which asks the CoT judge to critique and revise its initial prediction. All variants are evaluated on PlanCraft using the same coalitions, topologies, and three runs as exact removal.

\Cref{tab:judge_design_sensitivity} shows that none of these variants consistently improves both rank and absolute agreement across topologies. Self-consistency improves rank correlation in three of four topologies, but does not improve calibration and incurs \(3\times\) the judge cost. Rubric-based CoT, self-consistency, and self-refinement perform worse, reducing rank agreement and increasing upward bias in estimated coalition utility. On Centralized and Hybrid, they overestimate mean coalition value by $0.43$--$0.47$, despite true means of only $0.11$--$0.15$, further degrading the already negative identity-line $R^2$. For reference, independent exact-removal re-runs achieve $\rho=0.76$--$0.86$. These results suggest that standard judge-enhancement strategies do not resolve the limitations of introspective removal: additional reasoning or ensembling does not substitute for re-executing the coalition.

\begin{table}[t!]
    \centering
    \caption{
     Normalized Shapley attribution entropy ($H / \log_2 n$) across MAS architectures, with total evaluation token usage and cost (USD) over all $2^n-1$ coalitions, macro-averaged over architectures. Shaded cells indicate high influence concentration (entropy $< 0.6$).
    }
    \small
    \resizebox{0.75\columnwidth}{!}{%
    \begin{tabular}{lrrrr|rr}
    \toprule
    \textbf{Intervention Protocol} & \textbf{Independent.} & \textbf{Decentralized.}
    & \textbf{Centralized.} & \textbf{Hybrid} & \textbf{Tokens} & \textbf{Cost (\$)} \\
    \midrule
    \multicolumn{7}{l}{\textbf{PlanCraft}} \\
    \midrule
    Agent ablation         & 0.724 & 0.713 & \cellcolor{gray!25}0.580 & \cellcolor{gray!25}0.599 &  95.69M & 264.47 \\
    Introspective removal  & 0.860 & 0.808 & \cellcolor{gray!25}0.422 & \cellcolor{gray!25}0.424 &  20.52M & 6.70 \\
    Model replacement      & 0.711 & \cellcolor{gray!25}0.469 & 0.929 & 0.817 &  80.97M & 228.36 \\
    Prompt optimization & 0.796 & 0.657 & 0.811 & 0.728 & 366.65M & 708.51 \\
    \midrule
    \multicolumn{7}{l}{\textbf{WorkBench}} \\
    \midrule
    Agent ablation         & 0.976 & 0.865 & 0.739 & 0.670 &  21.65M & 41.20 \\
    Introspective removal  & 0.960 & 0.829 & 0.720 & \cellcolor{gray!25}0.579 &   9.23M & 3.70 \\
    Model replacement      & 0.815 & 0.962 & 0.933 & 0.939 &  32.76M & 54.61 \\
    Prompt optimization    & 0.803 & 0.948 & 0.755 & 0.822 & 108.22M & 142.21 \\
    \midrule
    \multicolumn{7}{l}{\textbf{BrowseComp-Plus}} \\
    \midrule
    Agent ablation         & 0.948 & 0.986 & 0.629 & \cellcolor{gray!25}0.581 & 537.88M & 1417.20 \\
    Introspective removal  & 0.986 & 0.996 & \cellcolor{gray!25}0.155 & \cellcolor{gray!25}0.132 &  17.99M & 5.49 \\
    Model replacement      & 0.928 & 0.912 & 0.827 & 0.895 & 586.60M & 1500.49 \\
    Prompt optimization    & -- & 0.934 & 0.857 & 0.698 & 826.53M & 1502.61 \\
    \midrule
    \multicolumn{7}{l}{\textbf{LiveCodeBench}} \\
    \midrule
    Agent ablation         & \cellcolor{gray!25}0.107 & \cellcolor{gray!25}0.148 & \cellcolor{gray!25}0.481 & \cellcolor{gray!25}0.483 & 118.47M & 436.97 \\
    Introspective removal  & \cellcolor{gray!25}0.217 & 0.816 & \cellcolor{gray!25}0.455 & \cellcolor{gray!25}0.499 &  17.48M & 5.80 \\
    Model replacement      & \cellcolor{gray!25}0.315 & 0.632 & 0.754 & 0.662 & 109.50M & 344.79 \\
    Prompt optimization    & 0.891 & 0.918 & 0.754 & 0.802 & 191.72M & 538.57 \\
    \bottomrule
    \end{tabular}
    }
    \label{tab:attribution-distribution}
\end{table}

 \begin{table}[t!]
      \centering
      \caption{
       Normalized LOO attribution entropy ($H / \log_2 n$) across MAS architectures, with total evaluation token usage and cost (USD) over the $n$ leave-one-out coalitions plus the full team, macro-averaged over architectures. Shaded cells indicate high influence concentration (entropy $< 0.6$).
      }
      \small
      \resizebox{0.75\columnwidth}{!}{%
      \begin{tabular}{lrrrr|rr}
      \toprule
      \textbf{Intervention Protocol} & \textbf{Independent.} & \textbf{Decentralized.}
      & \textbf{Centralized.} & \textbf{Hybrid} & \textbf{Tokens} & \textbf{Cost (\$)} \\
      \midrule
      \multicolumn{7}{l}{\textbf{PlanCraft}} \\
      \midrule
      Agent ablation         & \cellcolor{gray!25}0.457 & 0.815 & \cellcolor{gray!25}0.534 & 0.632 &  21.48M & 57.18 \\
      Introspective removal  & 0.717 & 0.906 & 0.751 & \cellcolor{gray!25}0.513 &   2.69M & 0.81 \\
      Model replacement      & \cellcolor{gray!25}0.590 & 0.623 & \cellcolor{gray!25}0.387 & 0.887 &  16.82M & 48.14 \\
      Prompt optimization    & 0.657 & 0.772 & 0.947 & 0.925 &  71.38M & 111.86 \\
      \midrule
      \multicolumn{7}{l}{\textbf{WorkBench}} \\
      \midrule
      Agent ablation         & 0.943 & 0.906 & 0.794 & \cellcolor{gray!25}0.459 &   5.81M & 11.34 \\
      Introspective removal  & 0.714 & 0.611 & 0.732 & 0.809 &   1.24M & 0.51 \\
      Model replacement      & 0.793 & 0.826 & 0.678 & 0.974 &   6.59M & 12.14 \\
      Prompt optimization    & 0.870 & 0.980 & 0.933 & 0.910 &  18.82M & 19.76 \\
      \midrule
      \multicolumn{7}{l}{\textbf{BrowseComp-Plus}} \\
      \midrule
      Agent ablation         & \cellcolor{gray!25}0.594 & 0.979 & \cellcolor{gray!25}0.518 & \cellcolor{gray!25}0.429 & 195.78M & 512.70 \\
      Introspective removal  & 0.778 & 0.896 & \cellcolor{gray!25}0.308 & \cellcolor{gray!25}0.221 &   4.18M & 1.17 \\
      Model replacement      & 0.792 & 0.765 & 0.984 & 0.960 & 199.67M & 509.64 \\
      Prompt optimization    & -- & 0.689 & 0.814 & 0.845 & 196.28M & 339.80 \\
      \midrule
      \multicolumn{7}{l}{\textbf{LiveCodeBench}} \\
      \midrule
      Agent ablation         & \cellcolor{gray!25}0.324 & \cellcolor{gray!25}0.348 & \cellcolor{gray!25}0.524 & \cellcolor{gray!25}0.495 &  45.30M & 168.95 \\
      Introspective removal  & \cellcolor{gray!25}0.251 & \cellcolor{gray!25}0.499 & \cellcolor{gray!25}0.360 & \cellcolor{gray!25}0.306 &   4.02M & 1.29 \\
      Model replacement      & 0.721 & 0.901 & 0.815 & 0.716 &  36.76M & 119.04 \\
      Prompt optimization    & 0.921 & 0.921 & 0.772 & 0.969 &  48.44M & 125.65 \\
      \bottomrule
      \end{tabular}
      }
      \label{tab:attribution-distribution-loo}
  \end{table}

  \begin{table}[t!]
    \centering
    \caption{
     Normalized Myerson-value attribution entropy ($H / \log_2 n$) across MAS architectures, with total evaluation token usage and cost (USD) over the coalitions connected in the communication graph (star with the orchestrator as hub for centralized, complete otherwise), macro-averaged over architectures. Shaded cells indicate high influence concentration (entropy $< 0.6$).
    }
    \small
    \resizebox{0.75\columnwidth}{!}{%
    \begin{tabular}{lrrrr|rr}
    \toprule
    \textbf{Intervention Protocol} & \textbf{Independent.} & \textbf{Decentralized.}
    & \textbf{Centralized.} & \textbf{Hybrid} & \textbf{Tokens} & \textbf{Cost (\$)} \\
    \midrule
    \multicolumn{7}{l}{\textbf{PlanCraft}} \\
    \midrule
    Agent ablation         & 0.724 & 0.713 & \cellcolor{gray!25}0.580 & \cellcolor{gray!25}0.599 &  95.69M & 264.47 \\
    Introspective removal  & 0.860 & 0.808 & \cellcolor{gray!25}0.422 & \cellcolor{gray!25}0.424 &  17.63M & 5.76 \\
    Model replacement      & 0.711 & \cellcolor{gray!25}0.469 & 0.805 & 0.817 &  72.31M & 202.87 \\
    Prompt optimization & 0.796 & 0.657 & 0.773 & 0.728 & 328.49M & 613.12 \\
    \midrule
    \multicolumn{7}{l}{\textbf{WorkBench}} \\
    \midrule
    Agent ablation         & 0.976 & 0.865 & 0.739 & 0.670 &  21.65M & 41.20 \\
    Introspective removal  & 0.960 & 0.829 & 0.720 & \cellcolor{gray!25}0.579 &   7.84M & 3.19 \\
    Model replacement      & 0.815 & 0.962 & 0.900 & 0.939 &  28.93M & 48.19 \\
    Prompt optimization    & 0.803 & 0.948 & 0.867 & 0.822 &  94.73M & 122.32 \\
    \midrule
    \multicolumn{7}{l}{\textbf{BrowseComp-Plus}} \\
    \midrule
    Agent ablation         & 0.948 & 0.986 & 0.629 & \cellcolor{gray!25}0.581 & 537.88M & 1417.20 \\
    Introspective removal  & 0.986 & 0.996 & \cellcolor{gray!25}0.155 & \cellcolor{gray!25}0.132 &  15.69M & 4.83 \\
    Model replacement      & 0.928 & 0.912 & 0.984 & 0.895 & 492.86M & 1259.85 \\
    Prompt optimization    & -- & 0.934 & 0.820 & 0.698 & 721.78M & 1249.53 \\
    \midrule
    \multicolumn{7}{l}{\textbf{LiveCodeBench}} \\
    \midrule
    Agent ablation         & \cellcolor{gray!25}0.107 & \cellcolor{gray!25}0.148 & \cellcolor{gray!25}0.481 & \cellcolor{gray!25}0.483 & 118.47M & 436.97 \\
    Introspective removal  & \cellcolor{gray!25}0.217 & 0.816 & \cellcolor{gray!25}0.455 & \cellcolor{gray!25}0.499 &  15.31M & 5.08 \\
    Model replacement      & \cellcolor{gray!25}0.315 & 0.632 & 0.847 & 0.662 &  96.75M & 304.49 \\
    Prompt optimization    & 0.891 & 0.918 & 0.655 & 0.802 & 170.63M & 476.00 \\
    \bottomrule
    \end{tabular}
    }
    \label{tab:attribution-distribution-myerson}
\end{table}

\begin{table}[t!]
    \centering
    \caption{
     Normalized Owen-value attribution entropy ($H / \log_2 n$) across MAS architectures, with total evaluation token usage and cost (USD) over all $2^n-1$ coalitions, macro-averaged over architectures. The coalition structure is $\{\texttt{workers}\},\{\texttt{orchestrator}\}$; without an orchestrator the Owen value equals the Shapley value. Shaded cells indicate high influence concentration (entropy $< 0.6$).
    }
    \small
    \resizebox{0.75\columnwidth}{!}{%
    \begin{tabular}{lrrrr|rr}
    \toprule
    \textbf{Intervention Protocol} & \textbf{Independent.} & \textbf{Decentralized.}
    & \textbf{Centralized.} & \textbf{Hybrid} & \textbf{Tokens} & \textbf{Cost (\$)} \\
    \midrule
    \multicolumn{7}{l}{\textbf{PlanCraft}} \\
    \midrule
    Agent ablation         & 0.724 & 0.713 & 0.696 & 0.705 &  95.69M & 264.47 \\
    Introspective removal  & 0.860 & 0.808 & \cellcolor{gray!25}0.498 & 0.620 &  20.52M & 6.70 \\
    Model replacement      & 0.711 & \cellcolor{gray!25}0.469 & 0.952 & 0.908 &  80.97M & 228.36 \\
    Prompt optimization & 0.796 & 0.657 & 0.712 & 0.632 & 366.65M & 708.51 \\
    \midrule
    \multicolumn{7}{l}{\textbf{WorkBench}} \\
    \midrule
    Agent ablation         & 0.976 & 0.865 & 0.811 & 0.828 &  21.65M & 41.20 \\
    Introspective removal  & 0.960 & 0.829 & 0.785 & 0.752 &   9.23M & 3.70 \\
    Model replacement      & 0.815 & 0.962 & 0.866 & 0.949 &  32.76M & 54.61 \\
    Prompt optimization    & 0.803 & 0.948 & 0.687 & 0.806 & 108.22M & 142.21 \\
    \midrule
    \multicolumn{7}{l}{\textbf{BrowseComp-Plus}} \\
    \midrule
    Agent ablation         & 0.948 & 0.986 & 0.860 & 0.859 & 537.88M & 1417.20 \\
    Introspective removal  & 0.986 & 0.996 & \cellcolor{gray!25}0.106 & \cellcolor{gray!25}0.134 &  17.99M & 5.49 \\
    Model replacement      & 0.928 & 0.912 & 0.956 & 0.902 & 586.60M & 1500.49 \\
    Prompt optimization    & -- & 0.934 & 0.669 & \cellcolor{gray!25}0.560 & 826.53M & 1502.61 \\
    \midrule
    \multicolumn{7}{l}{\textbf{LiveCodeBench}} \\
    \midrule
    Agent ablation         & \cellcolor{gray!25}0.107 & \cellcolor{gray!25}0.148 & \cellcolor{gray!25}0.479 & \cellcolor{gray!25}0.470 & 118.47M & 436.97 \\
    Introspective removal  & \cellcolor{gray!25}0.217 & 0.816 & 0.668 & 0.680 &  17.48M & 5.80 \\
    Model replacement      & \cellcolor{gray!25}0.315 & 0.632 & 0.836 & 0.692 & 109.50M & 344.79 \\
    Prompt optimization    & 0.891 & 0.918 & 0.715 & 0.702 & 191.72M & 538.57 \\
    \bottomrule
    \end{tabular}
    }
    \label{tab:attribution-distribution-owen}
\end{table}

\clearpage
\subsection{Agent-level effects of each intervention.}
\begin{figure}[h!]
    \centering
    \includegraphics[width=0.75\linewidth]{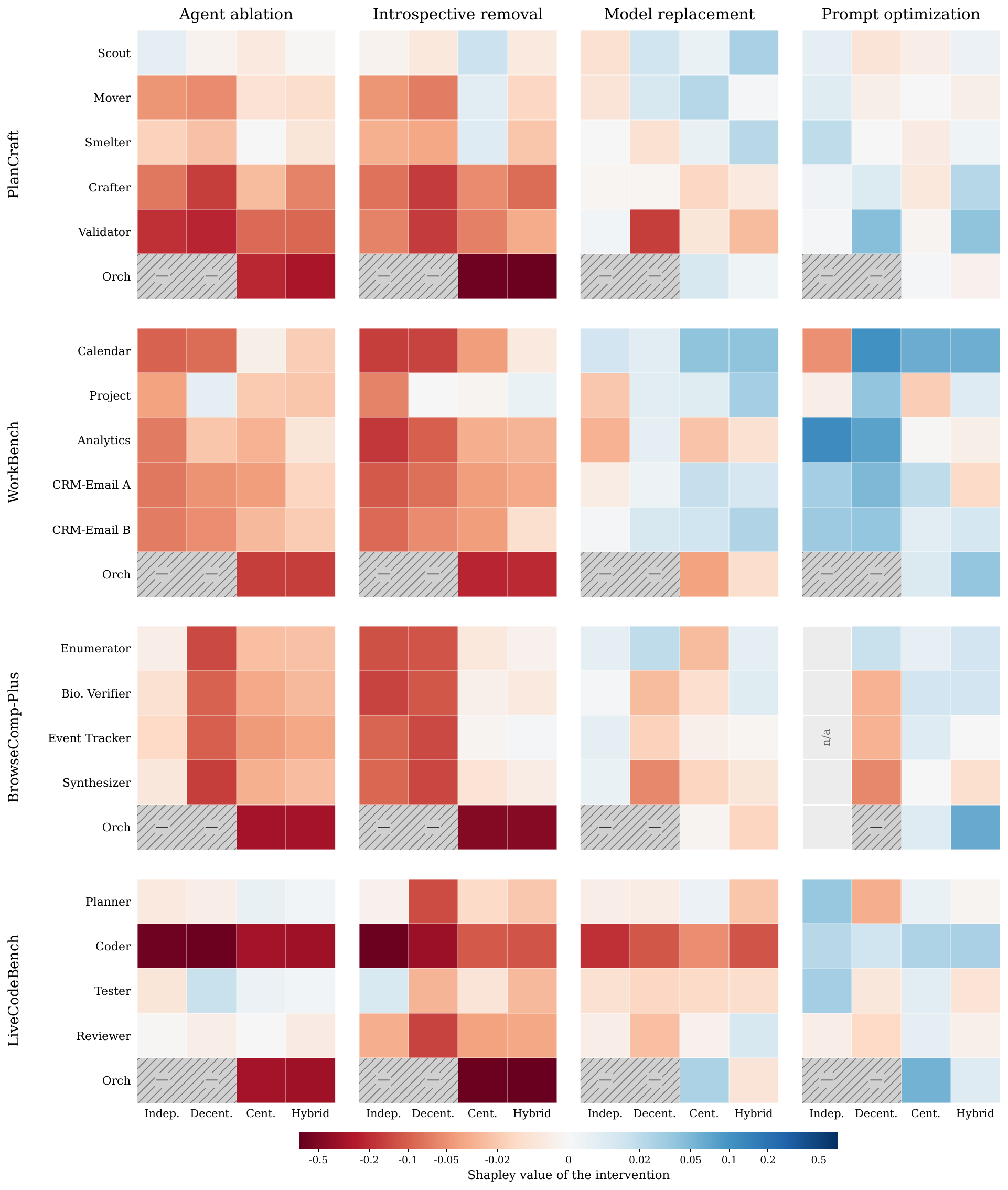}
    \caption{\textbf{Agent-level Shapley values across intervention protocols for GPT-5.4 teams.}
    Each cell shows the Shapley value of intervening on an agent (rows) under a communication topology (columns). Interventions are agent ablation, introspective removal, model replacement with GPT-OSS-120B, and prompt optimization. Blue (red) indicates a positive (negative) contribution to task success, using a shared symmetric logarithmic scale. Hatched cells denote agents absent from a topology, and n/a denotes unavailable prompt-optimization settings.}
    \label{fig:shapley_heatmap_4bench}
\end{figure}

\Cref{fig:shapley_heatmap_4bench} reports the Shapley value of intervening on each agent. The intervention protocols differ substantially in both the magnitude and sign of their effects.

\noindent\textbf{Removal.}
Agent ablation is predominantly harmful: removing an agent reduces success in $91\%$ of cells. Attribution is also concentrated in structurally important roles. The orchestrator has the most negative ablation attribution in every centralized and hybrid system ($-0.16$ to $-0.34$), while the LiveCodeBench Coder dominates among workers, with ablation values of $-0.62$ and $-0.64$ in the independent and decentralized topologies. Similarly, PlanCraft assigns most worker credit to the Validator and Crafter, which have access to the full action set. Only $7$ of $80$ ablation cells are positive, and these gains are small ($\leq 0.017$). Introspective removal broadly preserves this structure but produces larger-magnitude values overall (median absolute value $0.048$ vs.\ $0.026$ for ablation), particularly for orchestrators.

\noindent\textbf{Model replacement.}
Replacement produces substantially smaller effects (median absolute value $0.012$) and more mixed signs, with $46\%$ of cells positive. Most agents can therefore be replaced by GPT-OSS-120B with limited change in task success. Larger negative effects remain concentrated in capability-critical roles: replacing the LiveCodeBench Coder reduces success across all topologies, while replacing the PlanCraft Validator is also harmful in several settings. In contrast, replacing the orchestrator generally has little effect ($-0.04$ to $+0.03$).

\noindent\textbf{Prompt optimization.}
Prompt optimization is the only intervention with predominantly positive effects, with $61\%$ of cells improving task success. Gains are concentrated in particular agents, including the WorkBench Analytics and Calendar roles, while optimizing some agents decreases performance. Notably, the LiveCodeBench Coder---which dominates under ablation and replacement---gains only $0.02$--$0.03$ from prompt optimization. Thus, agents that are critical to current system performance are not necessarily those with the greatest optimization headroom.

The identity of high-impact agents depends on the intervention: removal emphasizes structural necessity, replacement reflects dependence on backbone capability, and prompt optimization captures room for behavioral improvement. 

\subsection{Rank agreement across Coalition Distribution}

\begin{figure}[h!]
    \centering
    \includegraphics[width=.9\linewidth]{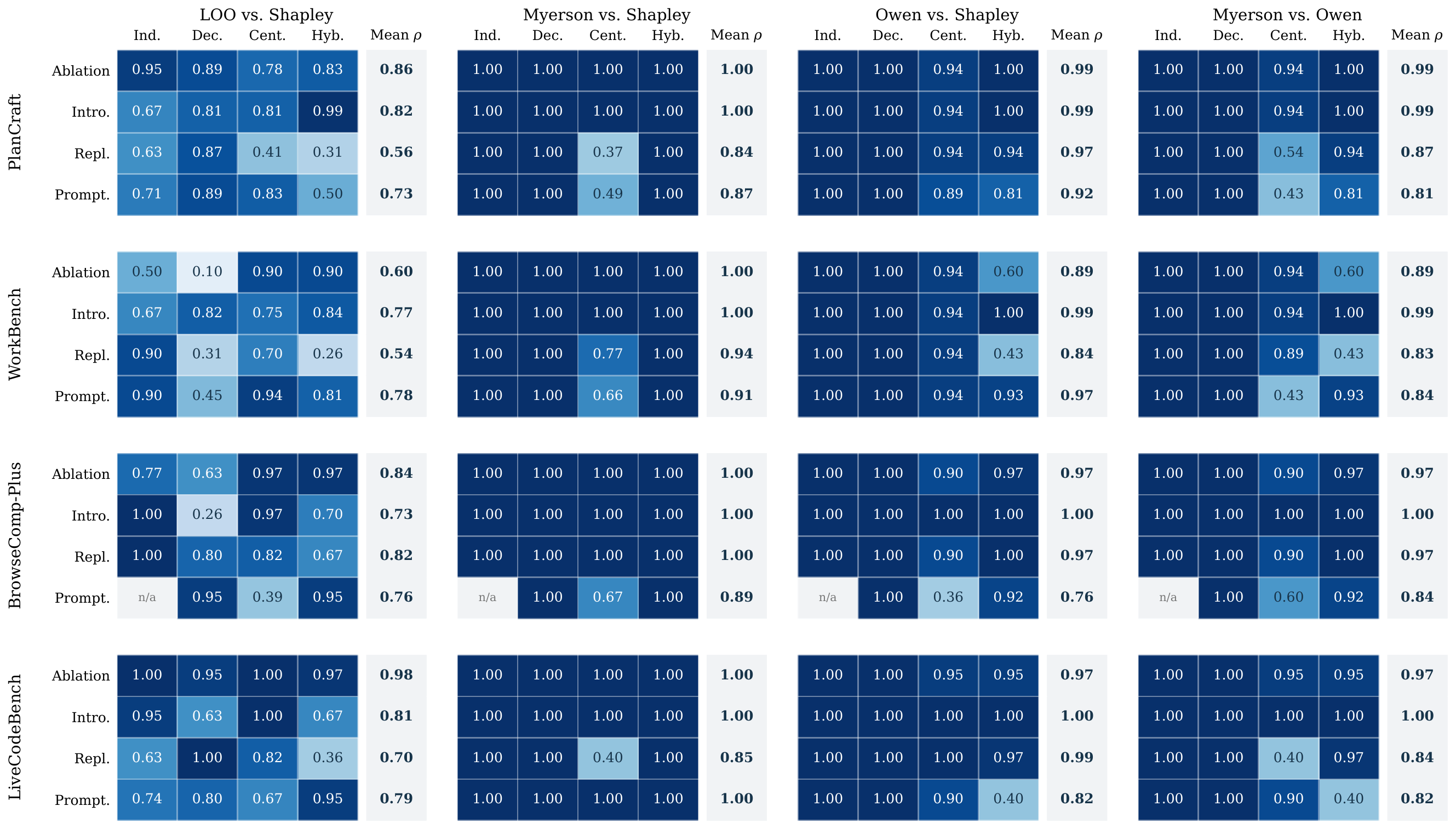}
    \caption{
    \textbf{Pairwise rank agreement across attribution methods.}
    Spearman correlation $\rho$ for LOO--Shapley, Myerson--Shapley,
    Owen--Shapley, and Myerson--Owen across datasets, intervention protocols, and communication topologies for GPT-5.4 teams. Ind., Dec., Cent., and Hyb.\ denote independent, decentralized, centralized, and hybrid topologies, respectively. The rightmost column reports the mean across available topologies. Cells marked n/a indicate settings without an independent prompt-optimization game.
    }
    \label{fig:rank_agreement_methods}
\end{figure}

\Cref{fig:rank_agreement_methods} compares the rankings produced by LOO, Myerson, and Owen against Shapley across datasets, intervention protocols, and communication topologies.

\noindent\textbf{LOO.}
LOO shows substantial agreement with Shapley, with mean correlations of $0.82$, $0.70$, $0.63$, and $0.77$ for ablation, introspective removal, replacement, and prompt optimization, respectively. Agreement is especially high for removal-based interventions in centralized and hybrid systems ($\rho=0.78$--$1.00$), consistent with these settings emphasizing the full-coalition context captured by LOO. Model replacement yields the lowest agreement overall.

\noindent\textbf{Myerson.}
Myerson coincides with Shapley whenever the communication graph is complete and also under centralized removal interventions. Differences therefore arise mainly for centralized replacement and prompt optimization, where correlations vary substantially across datasets.

\noindent\textbf{Owen.}
Owen generally agrees more closely with Shapley than LOO, with mean correlations of $0.91$, $0.89$, $0.79$, and $0.84$ across the four intervention protocols. It achieves higher mean agreement than LOO in $12$ of $16$ dataset--protocol pairs.

Despite differences in attribution values, the resulting agent rankings are generally stable across coalition distributions, with the strongest agreement among the combinatorial methods and substantial agreement from the cheaper LOO approximation.

\subsection{Rank-guided intervention utility on WorkBench, BrowseComp-Plus, and LiveCodeBench.}

\begin{table}[t!]
  \centering
  \caption{Per-run token usage and USD cost of each attribution method on PlanCraft, WorkBench, BrowseComp-Plus, and LiveCodeBench, averaged over $R=3$ runs and summed across all four architectures.}
  \resizebox{0.8\columnwidth}{!}{%
  \begin{tabular}{l c rr rr rr rr}
  \toprule
   & &
   \multicolumn{2}{c}{\textbf{Ablation}} &
   \multicolumn{2}{c}{\textbf{Introspective}} &
   \multicolumn{2}{c}{\textbf{Replacement}} &
   \multicolumn{2}{c}{\textbf{Prompt Opt.}} \\
  \cmidrule(lr){3-4}
  \cmidrule(lr){5-6}
  \cmidrule(lr){7-8}
  \cmidrule(lr){9-10}
  \textbf{Method} & \textbf{Coalitions}
  & \textbf{Tokens} & \textbf{Cost (\$)}
  & \textbf{Tokens} & \textbf{Cost (\$)}
  & \textbf{Tokens} & \textbf{Cost (\$)}
  & \textbf{Tokens} & \textbf{Cost (\$)} \\
  \midrule

  \multicolumn{10}{l}{\textbf{PlanCraft}} \\
  \midrule
  LOO & $n+1$
  & \cellcolor{green!20}80.9M & \cellcolor{green!20}213.40
  & \cellcolor{green!20}10.8M & \cellcolor{green!20}3.24
  & \cellcolor{green!20}111.9M & \cellcolor{green!20}192.57
  & \cellcolor{green!20}214.1M & \cellcolor{green!20}335.59 \\
  Shapley / Owen & $2^n-1$
  & 267.2M & 704.58
  & 82.1M & 26.82
  & 991.8M & 912.33
  & 1.35B & 2,719.31 \\
  Myerson$^\ddagger$ & $2^n-1$
  & 267.2M & 704.58
  & 70.5M & 23.03
  & 852.3M & 847.93
  & 1.20B & 2,337.74 \\

  \midrule
  \multicolumn{10}{l}{\textbf{WorkBench}} \\
  \midrule
  LOO & $n+1$
  & \cellcolor{green!20}22.5M & \cellcolor{green!20}44.14
  & \cellcolor{green!20}4.9M & \cellcolor{green!20}2.03
  & \cellcolor{green!20}37.7M & \cellcolor{green!20}53.29
  & \cellcolor{green!20}75.3M & \cellcolor{green!20}79.03 \\
  Shapley / Owen & $2^n-1$
  & 73.5M & 144.70
  & 36.9M & 14.79
  & 448.5M & 281.51
  & 432.9M & 568.85 \\
  Myerson$^\ddagger$ & $2^n-1$
  & 73.5M & 144.70
  & 31.4M & 12.74
  & 375.3M & 256.73
  & 378.9M & 489.27 \\

  \midrule
  \multicolumn{10}{l}{\textbf{BrowseComp-Plus}} \\
  \midrule
  LOO & $n+1$
  & \cellcolor{green!20}709.8M & \cellcolor{green!20}1,857.24
  & \cellcolor{green!20}16.7M & \cellcolor{green!20}4.66
  & \cellcolor{green!20}1.21B & \cellcolor{green!20}2,204.64
  & \cellcolor{green!20}588.8M & \cellcolor{green!20}1,019.41 \\
  Shapley / Owen & $2^n-1$
  & 1.58B & 4,164.96
  & 72.0M & 21.96
  & 6.37B & 6,717.42
  & 2.48B & 4,507.83 \\
  Myerson$^\ddagger$ & $2^n-1$
  & 1.58B & 4,164.96
  & 62.8M & 19.31
  & 5.44B & 5,560.83
  & 2.17B & 3,748.59 \\

  \midrule
  \multicolumn{10}{l}{\textbf{LiveCodeBench}} \\
  \midrule
  LOO & $n+1$
  & \cellcolor{green!20}165.1M & \cellcolor{green!20}624.98
  & \cellcolor{green!20}16.1M & \cellcolor{green!20}5.15
  & \cellcolor{green!20}260.4M & \cellcolor{green!20}716.21
  & \cellcolor{green!20}193.7M & \cellcolor{green!20}502.62 \\
  Shapley / Owen & $2^n-1$
  & 357.6M & 1,353.59
  & 69.9M & 23.20
  & 1.39B & 2,012.90
  & 766.9M & 2,154.29 \\
  Myerson$^\ddagger$ & $2^n-1$
  & 357.6M & 1,353.59
  & 61.3M & 20.33
  & 1.21B & 1,768.37
  & 682.5M & 1,904.01 \\
  \bottomrule
  \multicolumn{10}{l}{$^{\ddagger}$ Myerson uses fewer valid coalitions under the centralized topology.}
  \end{tabular}%
  }
  \label{tab:cost_tokens_merged}
\end{table}

\begin{figure}[h!]
    \centering
    \includegraphics[width=0.85\linewidth]{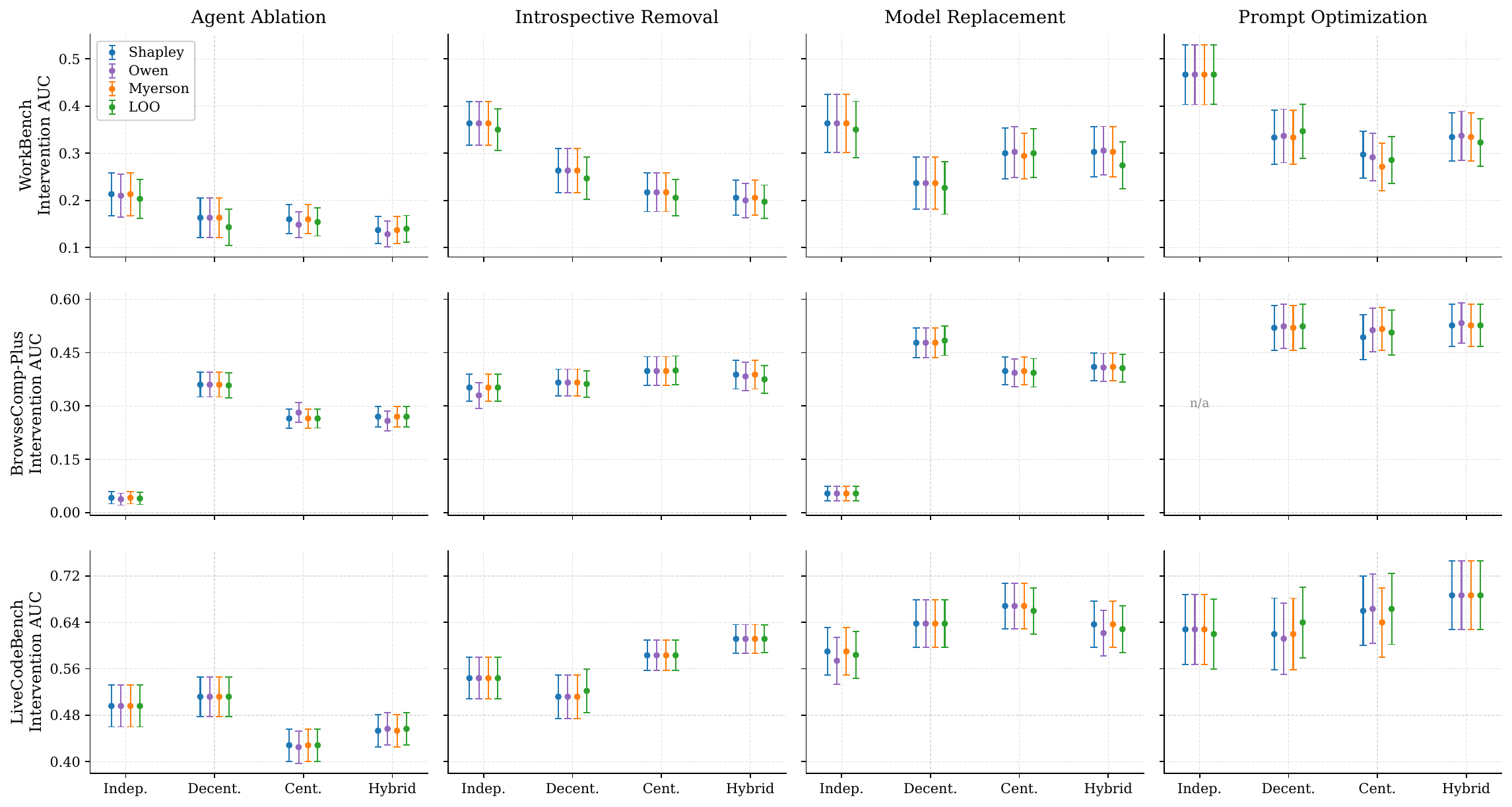}
    \caption{Comparison of intervention AUC across different intervention protocols (columns) and communication topologies ($x$-axis) in WorkBench, BrowseComp-Plus, and LiveCodeBench.}
    \label{fig:auc_workbench_browsecomp_livecodebench}
\end{figure}

\Cref{fig:auc_workbench_browsecomp_livecodebench} extends the analysis in \Cref{fig:auc_plancraft}(c) to the remaining benchmarks with GPT-5.4 teams. Agents are ranked by Shapley, Owen, Myerson, or LOO, and we evaluate the resulting top-$k$ interventions using AUC over $k=0,\dots,N$.

\begin{figure}
    \centering
    \includegraphics[width=1.0\linewidth]{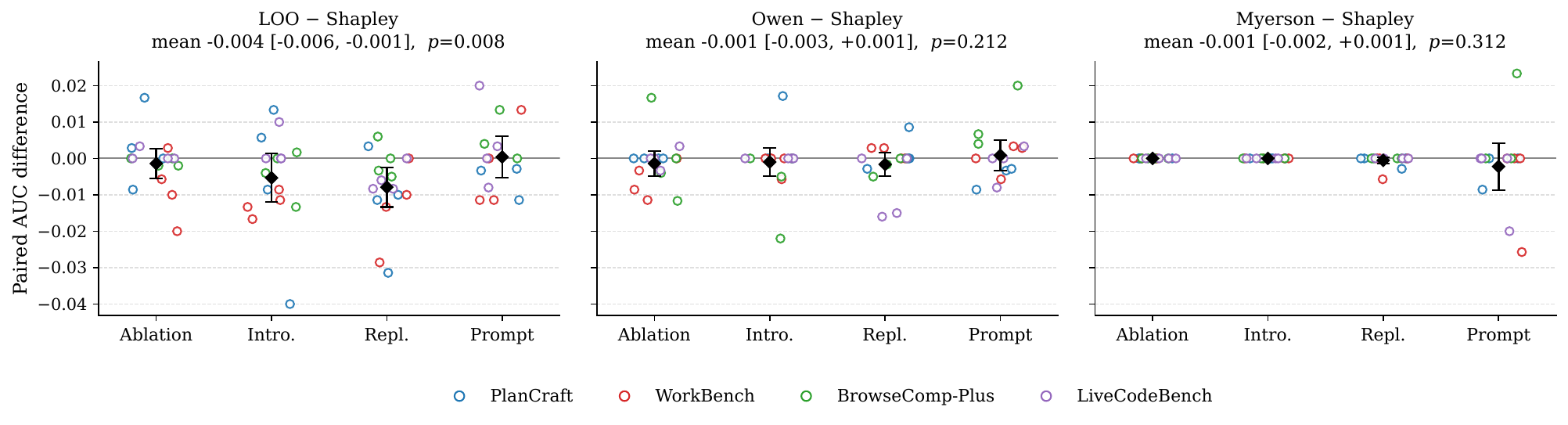}
    \caption{\textbf{Paired intervention-AUC differences from Shapley.} Each dot shows the AUC difference between LOO, Owen, or Myerson and Shapley for one benchmark--protocol--topology setting. Black diamonds show protocol means with 95\% confidence intervals; panel titles report the overall mean difference and Wilcoxon signed-rank $p$-value.}
    \label{fig:auc_paired_diff_gpt54}
\end{figure}

The four ranking methods yield nearly the same downstream intervention utility. To compare methods, we compute paired AUC differences from Shapley within each benchmark--protocol--topology setting (\Cref{fig:auc_paired_diff_gpt54}). LOO trails Shapley by only $0.004$ AUC on average (95\% CI $[0.001, 0.006]$; Wilcoxon signed-rank $p=0.008$), a small difference relative to a typical AUC of $0.38$, and the maximum per-setting difference is $0.040$. Owen and Myerson differ from Shapley by approximately $0.001$ AUC on average, with confidence intervals containing zero; Myerson yields an intervention curve identical to Shapley's in $90\%$ of settings. Thus, although the small LOO--Shapley difference is statistically detectable in aggregate, LOO recovers nearly all of Shapley's downstream intervention utility while requiring only $n+1$ rather than $2^n$ coalition evaluations (\Cref{tab:cost_tokens_merged}).

The differences across ranking methods are small relative to variation induced by the intervention protocol and topology. Within a setting, the four ranking methods span a median of only $0.009$ AUC (maximum $0.040$), whereas across protocols and topologies, Shapley AUC ranges from $0.26$ to $0.49$ within the same benchmark. This further indicates that downstream intervention utility is substantially more sensitive to the intervention game than to the choice of coalition distribution.

Agent ablation yields the lowest AUC on average ($0.29$), as removing essential agents can sharply reduce utility, whereas model replacement ($0.42$) and prompt optimization ($0.49$) preserve agent roles and achieve higher values; this ordering holds across all benchmarks. BrowseComp-Plus independent is an outlier under ablation and replacement (AUC $\approx0.05$), reflecting its low full-system utility ($4\%$ success versus $42\%$--$49\%$ under the other topologies). Introspective removal yields higher AUC than direct ablation in 14 of 16 benchmark--topology pairs (by $0.11$ on average), but its coalition utilities are judge-estimated rather than measured from the intervened MAS (\Cref{tab:coalition-agreement-combined}) and therefore characterize a different intervention game.

\subsection{Intervention outcomes by communication topology}
\begin{figure}[h!]
    \centering
    \includegraphics[width=0.85\linewidth]{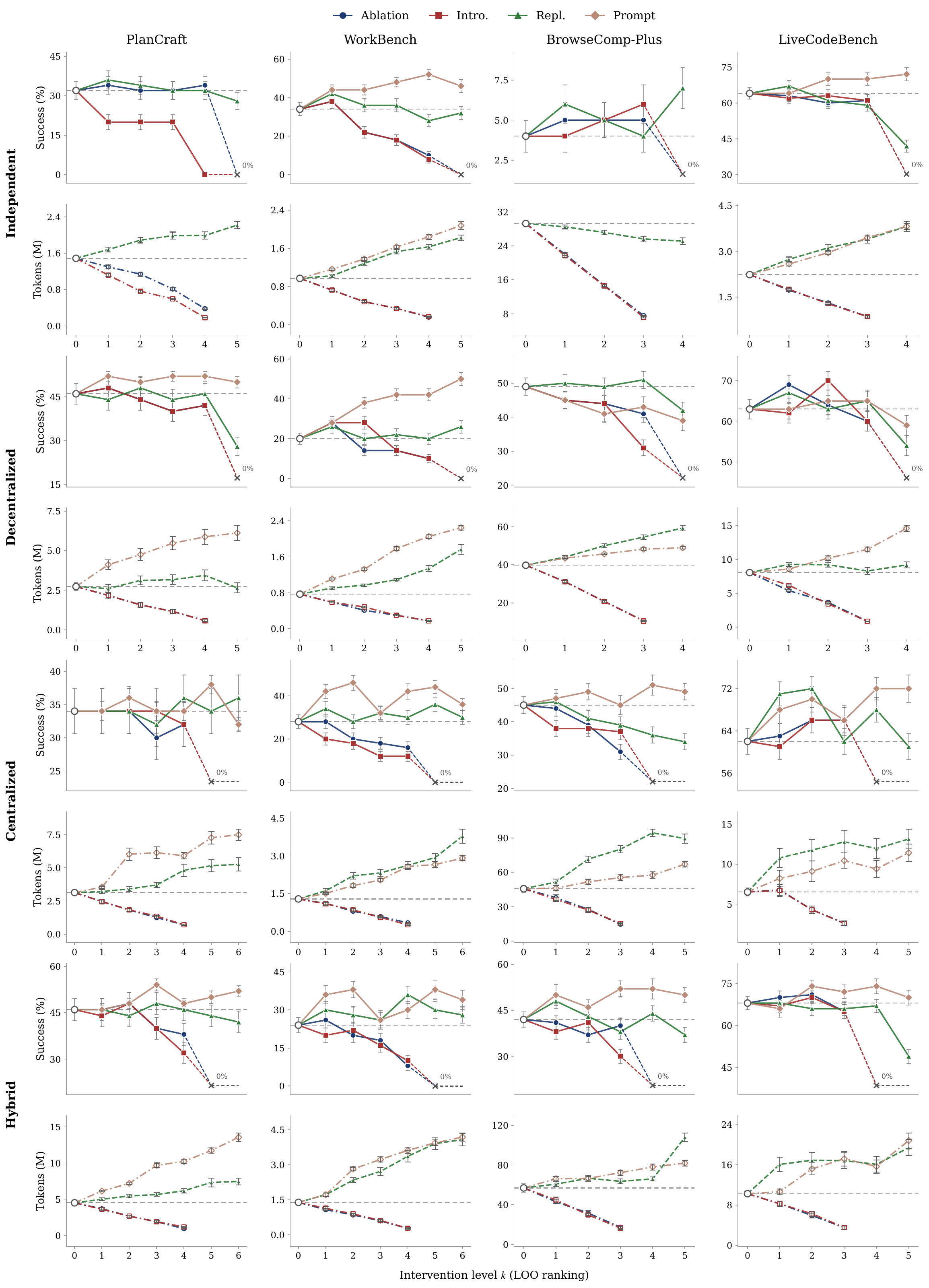}
    \caption{\textbf{LOO-guided interventions.}
    Columns denote datasets; row pairs show success (top) and token usage (bottom) across communication topologies. Hollow markers denote \(k=0\) controls; gray crosses indicate no-worker settings with \(0\%\) success. Error bars are 95\% confidence intervals.}
    \label{fig:intervention_4datasets}
\end{figure}

\Cref{fig:intervention_4datasets} reports the topology-specific results.  Overall trends remain similar, but effect sizes vary across datasets and topologies.

\textbf{Agent ablation.}
Agent ablation at \(k=1\) improves or preserves success in \(12/16\) dataset--topology settings while reducing token usage by up to \(33\%\). The largest gains occur on decentralized WorkBench (\(+8\%\)) and LiveCodeBench (\(+6\%\)), whereas BrowseComp-Plus decreases by \(1\)--\(4\%\) across communicating topologies. At larger \(k\), token reductions reach \(60\%\)--\(89\%\), but performance degradation is more pronounced on WorkBench (\(10\)--\(24\%\)) than on PlanCraft (at most \(8\%\)) or LiveCodeBench (at most \(3\%\)). Introspective removal is less stable at \(k=1\), decreasing success in \(11/16\) settings.

\textbf{Model replacement.}
Replacing the highest-ranked agent improves or preserves success in \(15/16\) settings, with gains up to \(9\)\% on centralized LiveCodeBench and \(6\)--\(8\)\% on WorkBench. Performance decreases at larger replacement depths, including \(18\)--\(22\)\% drops when all agents are replaced in decentralized PlanCraft and hybrid LiveCodeBench. Token usage increases in nearly all settings, reaching \(49\%\)--\(193\%\) above baseline on PlanCraft and WorkBench at maximum depth.

\textbf{Prompt optimization.}
Prompt optimization yields larger but more topology-dependent gains. On WorkBench, success improves by \(18\)--\(30\)\% across independent, decentralized, and centralized topologies; decentralized success increases from \(20\%\) to \(50\%\) when all agents are optimized. On LiveCodeBench, gains reach \(8\)--\(10\)\% in independent and centralized systems but remain within \(2\)\% in decentralized systems. On BrowseComp-Plus, centralized and hybrid systems improve by \(6\) and \(10\)\%, respectively. Peak success occurs before optimizing all agents in \(8/14\) settings, while token usage increases by \(23\%\)--\(201\%\) at the largest \(k\).

\subsection{Prompt-target selection across datasets and topologies}

Ranking agents under the prompt-optimization game selects more effective optimization targets than ranking them by ablation Shapley, as in CG-OPO. Averaged over \(k=1\text{--}3\), Prompt LOO improves over CG-OPO by \(2\text{--}6\) percentage points on every benchmark, with the largest gains on WorkBench. There, optimizing only the top-ranked agent improves success by \(11\) points, compared with \(4\) points under CG-OPO, while optimizing the top two recovers \(15\) of the \(18\) points achieved by optimizing all six agents. The rankings also often select different optimization targets, indicating that importance under removal does not necessarily reflect prompt-improvement potential.

Prompt LOO closely matches Prompt Shapley, differing by less than one point on average at every \(k\), while requiring \(4\text{--}6\times\) fewer attribution tokens (\(18\)M--\(168\)M versus \(104\)M--\(725\)M per benchmark) by evaluating only \(n+1\) rather than \(2^n\) coalitions. Shapley cost also grows sharply with topology, from \(26\)M tokens for independent LiveCodeBench to \(289\)M for its hybrid topology, and is comparable to the cost of CG-OPO's ablation Shapley.

\begin{figure}[h!]
\centering
\includegraphics[width=1.0\linewidth]{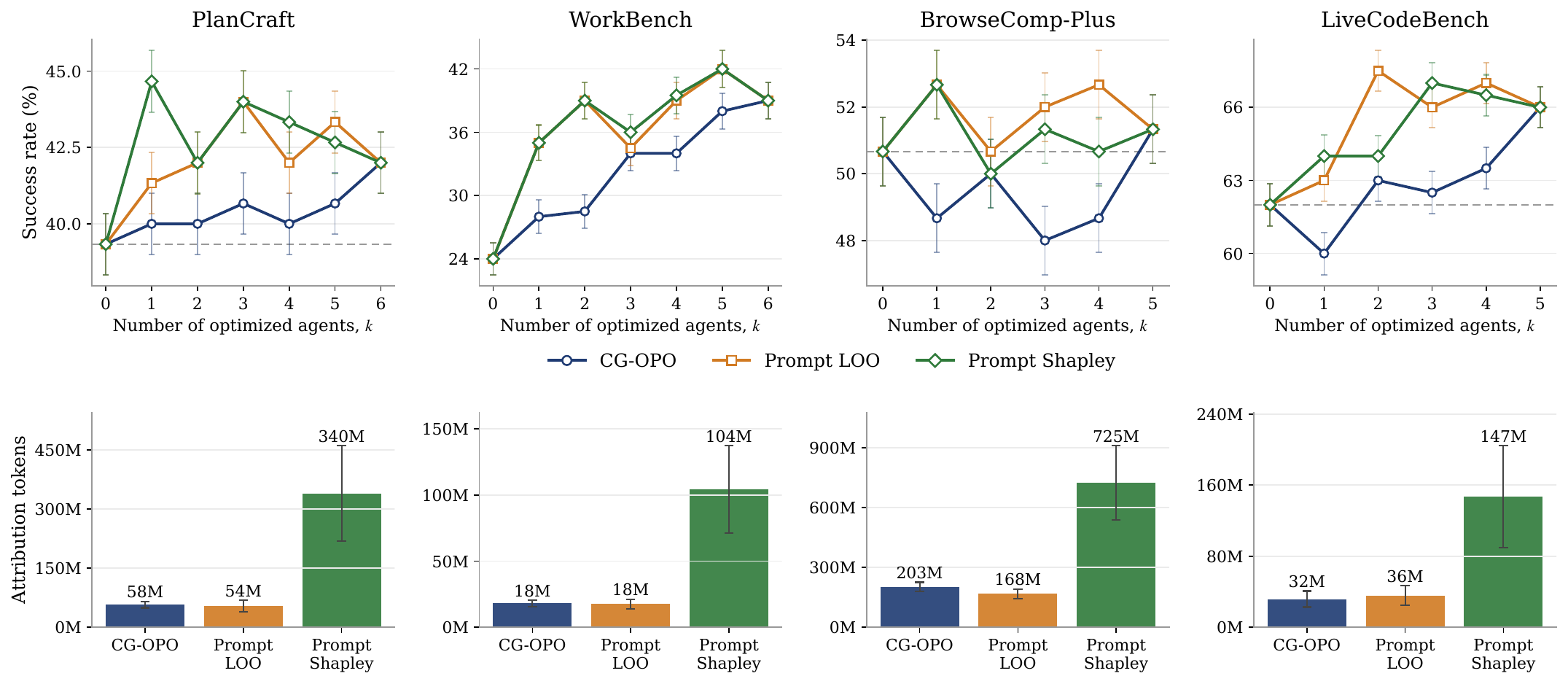}
\caption{\textbf{Prompt-target selection by dataset and topology.} Success as a function of the number of agents $k$ receiving optimized prompts under each ranking, with $k=0$ denoting the original-prompt team (dashed line). Panels denote benchmarks, with results averaged across communication topologies; BrowseComp-Plus have no independent prompt game. Error bars show 95\% confidence intervals.}
\label{fig:promptopt_ranking_comparison_gpt54_per_dataset_frac_std}
\end{figure}

The per-dataset results in \Cref{fig:promptopt_ranking_comparison_gpt54_per_dataset_frac_std} show that this advantage is consistent across benchmarks and their communication topologies. Prompt LOO yields the clearest gains over CG-OPO on WorkBench and PlanCraft, while BrowseComp-Plus shows smaller and less consistent improvements, consistent with its weaker overall response to prompt optimization. Across datasets, Prompt LOO remains close to Prompt Shapley while providing substantially lower attribution cost.

\clearpage
\subsection{Additional Results with Alternative Backbones}\label{sec:alternative_backbones}

We further evaluate whether our findings generalize across backbone models. We retain the same role-specialized team designs and communication topologies while varying the backbone by benchmark: PlanCraft uses five Gemini-2.5-Flash-Lite agents \citep{comanici2025gemini}, WorkBench uses five DeepSeek-V4-Flash agents \citep{deepseekai2026deepseekv4}, and BrowseComp-Plus uses four GPT-5-mini agents \citep{singh2025openai}. For model replacement, the target agent is replaced with Llama-3.1-8B on PlanCraft and WorkBench and Qwen3-30B on BrowseComp-Plus. For introspective removal, we use GPT-5-mini as the judge. Each configuration is evaluated over three independent runs using the same settings described in \Cref{apx:datasets}.

\subsubsection{Rank Agreement Across Coalition Distributions}

\begin{figure}[h!]
    \centering
    \includegraphics[width=1.0\linewidth]{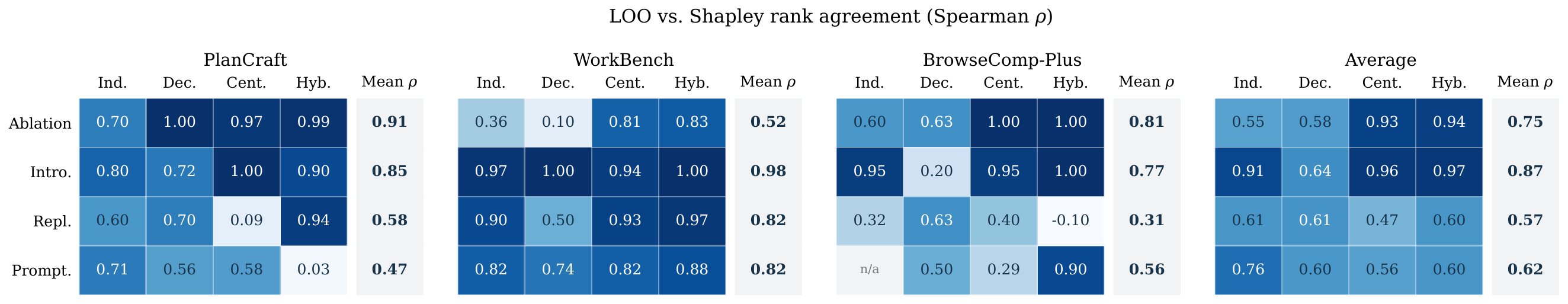}
      \caption{\textbf{LOO--Shapley rank agreement under alternative backbones.} Spearman $\rho$ between LOO and Shapley agent rankings across intervention protocols and communication topologies on PlanCraft, WorkBench, and BrowseComp-Plus. The final column reports the average across topologies; n/a denotes an unevaluated setting.}
        \label{fig:mixed-rank-agreement}
\end{figure}

\Cref{fig:mixed-rank-agreement} reports the Spearman correlation between LOO and Shapley rankings under alternative backbones. The pattern observed with GPT-5.4 teams largely persists. Averaged across the three benchmarks, agreement is higher for agent ablation ($\rho=0.75$) and introspective removal ($0.87$) than for model replacement ($0.57$) and prompt optimization ($0.62$), compared with $0.82$, $0.70$, $0.62$, and $0.77$, respectively, under GPT-5.4.

For the removal protocols, agreement is highest in centralized and hybrid topologies ($\rho=0.95$ on average), where both methods consistently identify the orchestrator as the most removal-sensitive agent, and lower in independent and decentralized systems ($0.67$). As with GPT-5.4, the lowest agent-ablation agreement occurs on WorkBench in the independent ($0.36$) and decentralized ($0.10$) topologies, where full-team utility saturates and worker contributions become difficult to distinguish. Model replacement and prompt optimization yield smaller differences among agents and consequently less stable full rankings. Nevertheless, LOO and Shapley identify the same top-ranked agent in 39 of 47 settings. Owen and Myerson rankings also remain close to Shapley, with average correlations of $\rho=0.85$ and $0.93$, respectively.

\subsubsection{Rank-Guided Intervention AUC}

\begin{figure}[h!]
    \centering
    \includegraphics[width=0.9\linewidth]{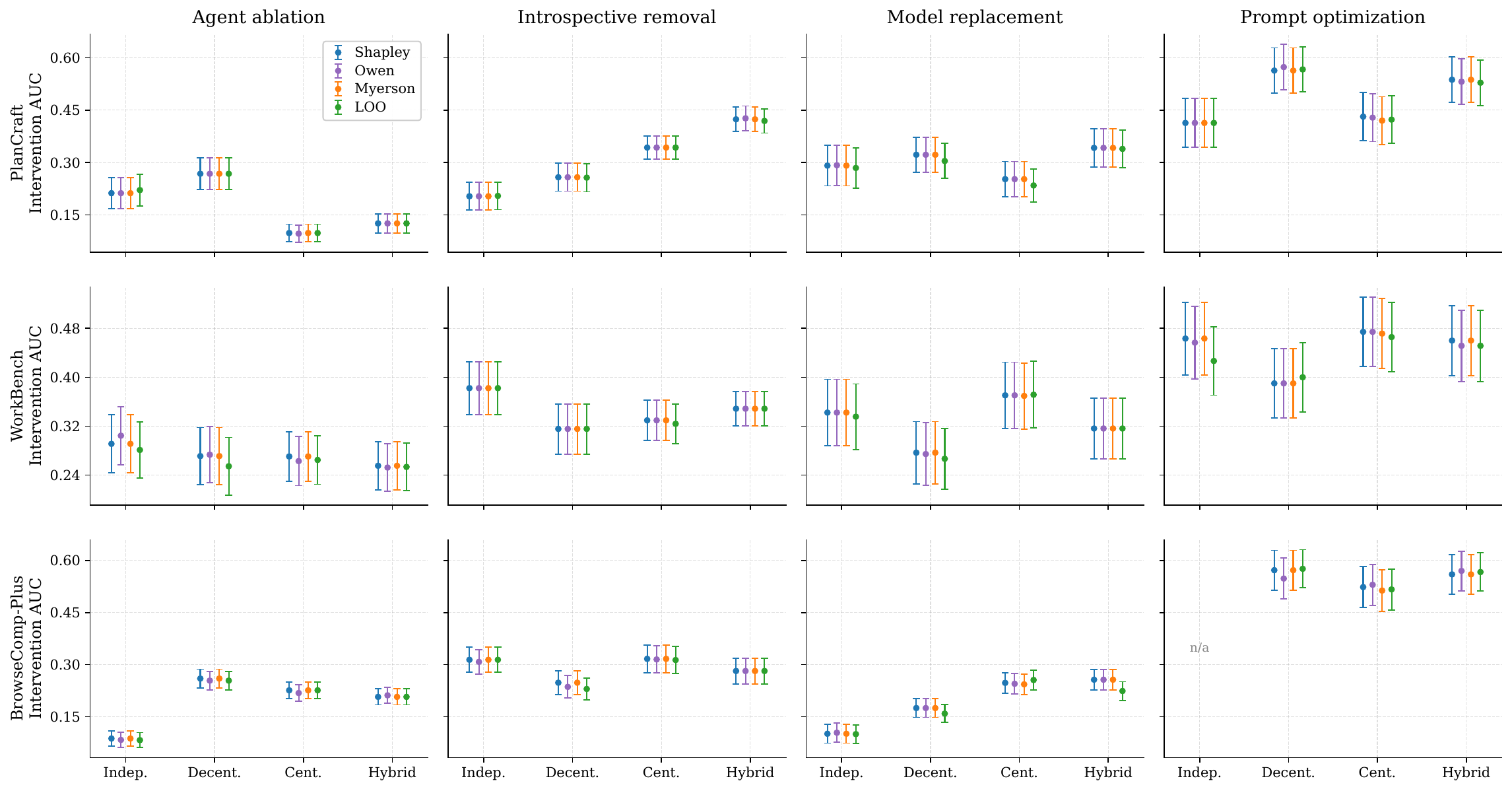}
    \caption{\textbf{Rank-guided intervention AUC under alternative backbones.} Rows denote benchmarks and columns denote intervention protocols. Within each panel, agents are ranked by Shapley, Owen, Myerson, or LOO for each topology. AUC is the mean utility across intervention depths $k=0,\dots,N$, using top-$k$ agents. Error bars show 95\% confidence intervals over evaluation tasks. n/a denotes an unevaluated setting.}
    \label{fig:mixed-auc}
\end{figure}

\begin{figure}[h!]
    \centering
    \includegraphics[width=1.0\linewidth]{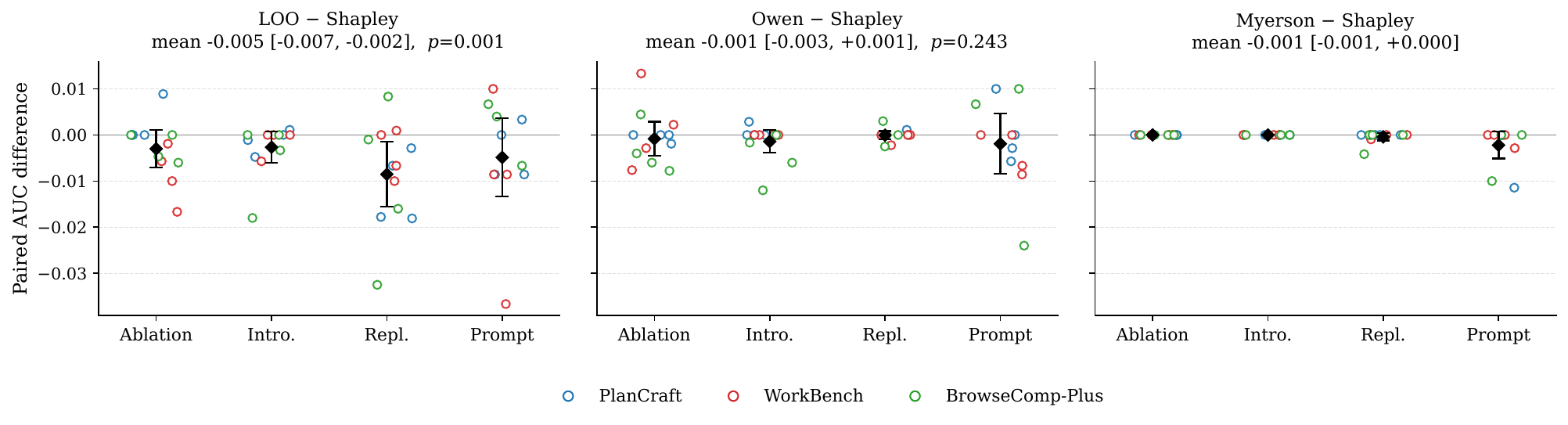}
    \caption{\textbf{Paired intervention-AUC differences from Shapley under alternative backbones.} Each dot shows the AUC difference between LOO, Owen, or Myerson and Shapley for one benchmark--protocol--topology setting. Black diamonds show protocol means with 95\% confidence intervals; panel titles report the overall mean difference and Wilcoxon signed-rank $p$-value.}
\label{fig:auc_paired_diff_mixed_models}
\end{figure}

Figure~\ref{fig:mixed-auc} compares intervention AUC across attribution rankings. Consistent with the GPT-5.4 results, differences in rank agreement translate into only small differences in downstream intervention utility. Across the 47 alternative-backbone settings, LOO trails Shapley by $0.005$ AUC on average (95\% CI $[0.002, 0.007]$; Wilcoxon signed-rank $p=0.001$), comparable to the $0.004$ gap observed with GPT-5.4 teams (95\% CI $[0.001, 0.006]$; $p=0.008$). The per-setting gap is at most $0.037$ and exceeds two task-sampling standard errors in only 4 of 47 settings, all under model replacement or prompt optimization and favoring Shapley. Owen and Myerson remain closer still, differing from Shapley by approximately $0.001$ AUC on average, with confidence intervals containing zero; Myerson exactly reproduces the Shapley intervention curve in 42 of 47 settings (\Cref{fig:auc_paired_diff_mixed_models}).

As with GPT-5.4 teams, variation across intervention protocols and topologies is substantially larger than variation across ranking methods. Within a setting, the four rankings span a median of only $0.008$ AUC (maximum $0.037$), whereas across protocols and topologies, Shapley AUC ranges from $0.22$ to $0.49$ within a benchmark. Agent ablation yields the lowest AUC on average ($0.21$), followed by model replacement ($0.27$) and prompt optimization ($0.49$). Replacement exceeds ablation on PlanCraft and WorkBench but not on BrowseComp-Plus, where replacing agents with Qwen3-30B provides little benefit over removing them. BrowseComp-Plus independent again has the lowest AUC under ablation and replacement ($0.09$ and $0.10$), reflecting its low full-system utility ($10\%$ success versus $22\%$--$29\%$ under the other topologies). PlanCraft centralized and hybrid ablation are similarly low ($0.10$ and $0.13$), as their single-worker coalitions achieve zero success. Introspective removal yields higher AUC than direct ablation in three quarters of benchmark--topology pairs, by $0.10$ on average, but its coalition utilities are judge-estimated rather than measured from the intervened MAS and therefore characterize a different intervention game.

\subsubsection{Intervention Outcomes by Communication Topology}

\begin{figure}[h!]
\centering
\includegraphics[width=0.8\linewidth]{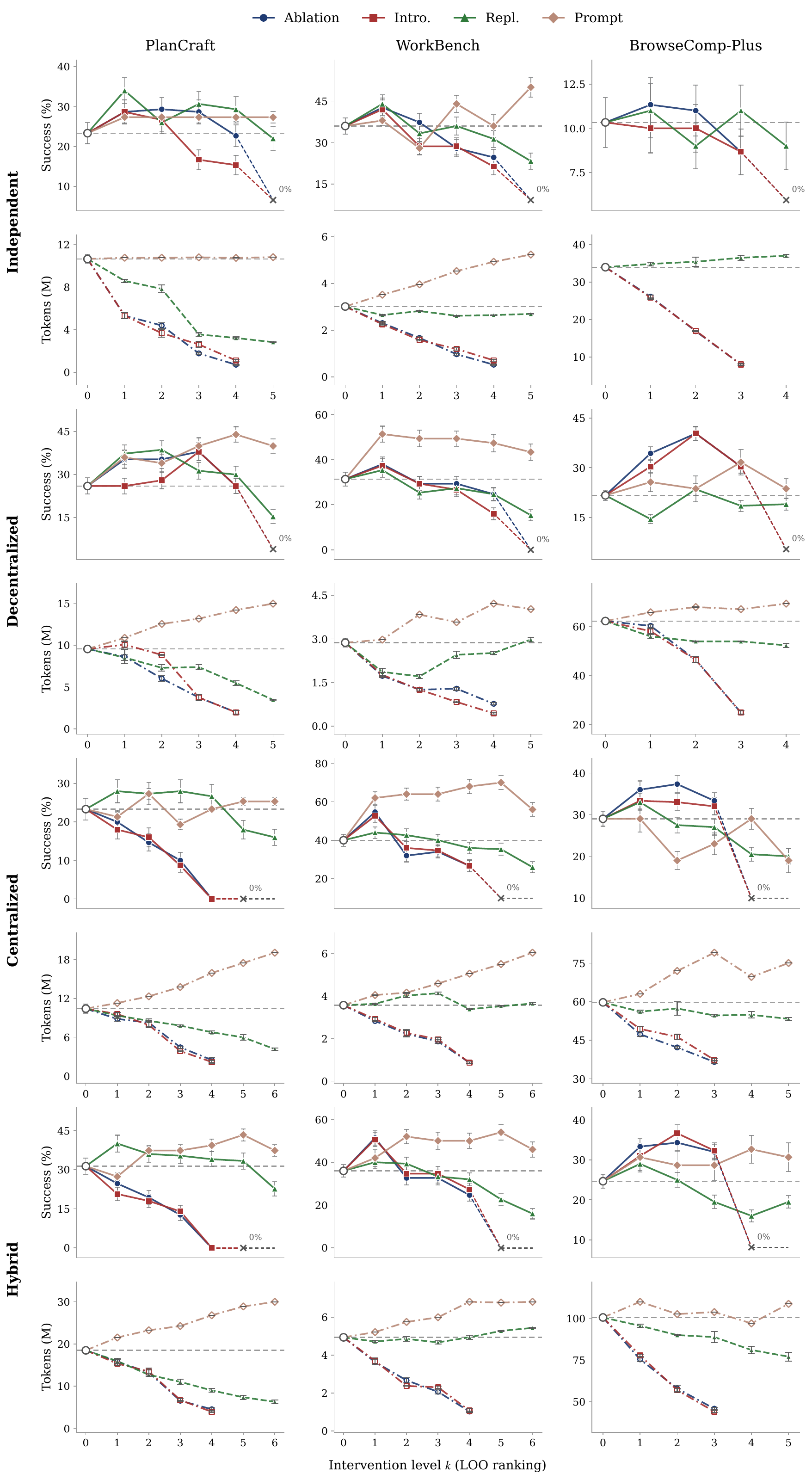}
\caption{\textbf{Intervention outcomes by topology under alternative backbones.} Success rate (top) and token usage (bottom) across intervention depths $k$ under LOO ranking. $k=0$ denotes the original system. Points are averaged over seeds; error bars show task-level uncertainty. Gray crosses denote configurations with defined zero utility when no worker or orchestrator remains.}
\label{fig:mixed-per-topology}
\end{figure}

Figure~\ref{fig:mixed-per-topology} shows success and token usage across intervention depths under LOO ranking. The qualitative trends from the GPT-5.4 experiments persist across alternative backbones. 

\textbf{Agent ablation} improves success after removing the highest-ranked agent in $83\%$ of benchmark--topology settings, by $6.4$ points on average, while usually reducing token usage. The largest gains occur on WorkBench centralized and hybrid ($+14.7$ and $+15.3$ points) and BrowseComp-Plus decentralized ($+12.6$). PlanCraft centralized and hybrid are the main exceptions, where success decreases monotonically as agents are removed. Deeper removal eventually degrades performance across all settings while continuing to reduce tokens.

\textbf{Introspective removal} generally produces similar removal orders and intervention curves to direct ablation, with the largest discrepancy on PlanCraft independent. \textbf{Model replacement} preserves or improves success after replacing the highest-ranked agent in $92\%$ of settings, but replacing all agents reduces performance throughout; in centralized and hybrid systems, performance is retained longer when the orchestrator keeps its original backbone.

\textbf{Prompt optimization} improves over the original-prompt control in $91\%$ of evaluated settings, with gains of up to $16$ points on WorkBench centralized, although the optimum often occurs before all agents are optimized. These gains consistently come with higher token usage; for example, PlanCraft centralized increases from $10.4$M to $19.1$M tokens. BrowseComp-Plus centralized is the primary exception, where optimization decreases success by $10$ points. Overall, the same performance--cost trade-offs observed with GPT-5.4 remain under alternative backbones, with orchestrator-based topologies particularly sensitive to removal.

\subsection{Sensitivity to the replacement model}\label{sec:model_sensitivity}
\begin{figure}[h!]
    \centering
    \includegraphics[width=1.0\linewidth]{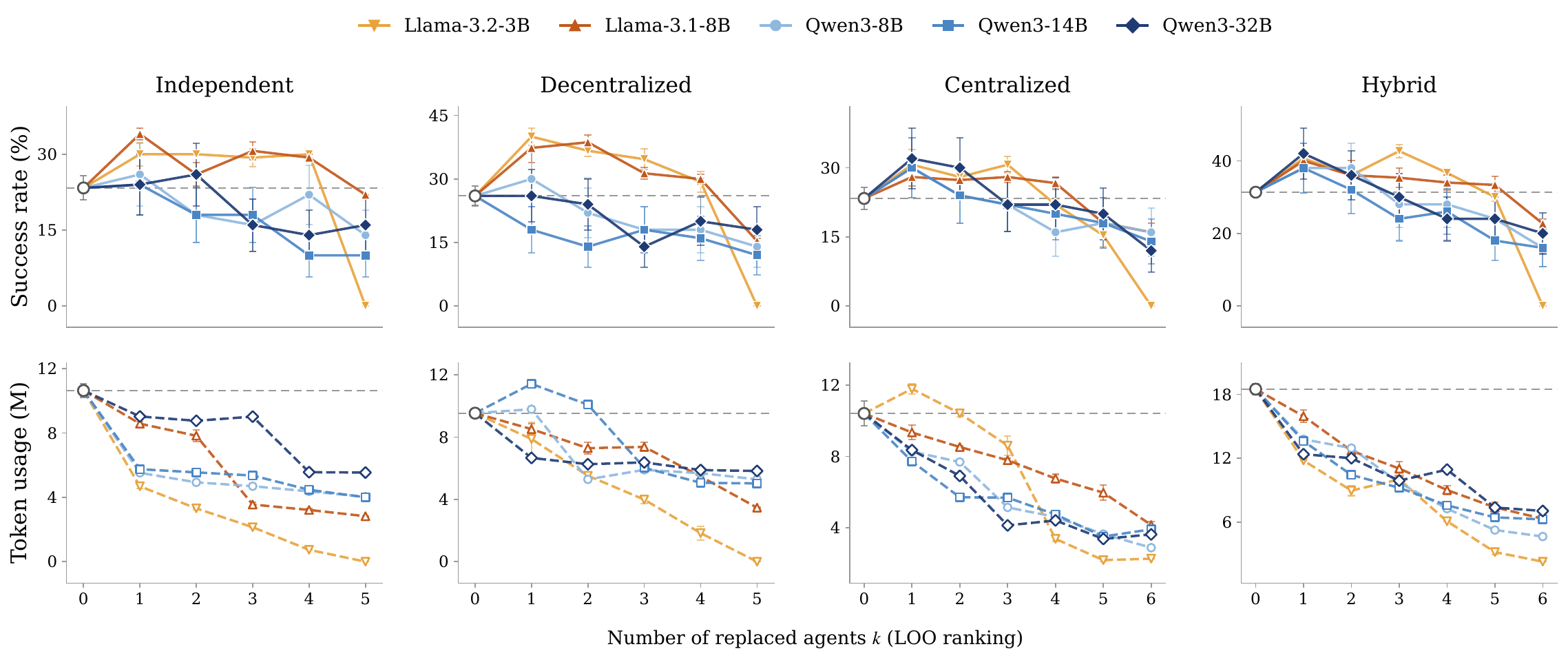}
    \caption{\textbf{LOO-guided interventions across communication topologies and replaced models.}
    Columns denote communication topologies; row pairs show success (top) and token usage (bottom). Hollow markers denote \(k=0\) controls; gray crosses indicate no-worker settings with \(0\%\) success. Error bars are 95\% confidence intervals.}
    \label{fig:replacement_sensitivity}
\end{figure}

Here, we test whether model-replacement results are sensitive to the choice of substitute model. We repeat LOO-guided replacement on PlanCraft using five substitute models from two model families, spanning \(3\text{B}\)--\(32\text{B}\) parameters: Llama-3.2-3B, Llama-3.1-8B, Qwen3-8B, Qwen3-14B, and Qwen3-32B (\Cref{fig:replacement_sensitivity}). In all settings, Gemini-2.5-Flash-Lite serves as the original backbone, and we recompute the LOO ranking separately for each replacement model. Shallow replacement is relatively robust across functioning substitutes, with \(k=1\) changing average success by only a few points while reducing token usage by \(14\%\text{--}25\%\). At larger \(k\), outcomes diverge: Llama-3.1-8B remains near baseline through \(k=3\),  whereas the Qwen3 models fall \(5\)--\(6\) points below it. Within the Qwen3 family, the 8B, 14B, and 32B variants perform similarly across most intervention depths, indicating that replacement quality does not scale monotonically with parameter count.

Cost and attribution are also substitute-dependent. Excluding Llama-3.2-3B, token usage decreases by \(39\%\text{--}48\%\) at \(k=3\) and \(53\%\text{--}66\%\) under full replacement, contrasting with GPT-OSS-120B in our GPT-5.4 experiments, where replacement increases token usage. LOO rankings likewise vary across substitutes, particularly among low-ranked agents. Thus, model-replacement attribution measures importance relative to a particular substitute, and neither performance nor cost is determined by parameter count alone. Llama-3.2-3B is an extreme case: replaced agents produce effectively no useful actions, causing replacement to behave more like agent ablation.

\clearpage
\onecolumn

\section{Prompt Templates}\label{apx:prompt_template}

\subsection{MAS System Prompt}

\begin{promptbox}{Lead orchestrator prompt}
\small
\textbf{System.}\ You are an intelligent lead agent who can orchestrate multi-agent
research tasks.

\medskip
\textbf{Task setup (user).}\ \PH{task description \& tools}.\ \PH{task behavior}.\
You coordinate a multi-agent team. Responsibilities: (1) plan and decompose the query,
(2) guide and give feedback to each agent, (3) decide when enough has been gathered,
and (4) synthesize a final answer. Task: \PH{instance}.

\medskip
\textbf{Planning.}\ Create exactly \PH{N} distinct subtasks. Return a JSON object with
a \texttt{subtasks} list, where each item contains \texttt{agent\_id},
\texttt{objective}, and \texttt{focus}, and a top-level \texttt{reasoning} field.

\medskip
\textbf{Per-agent coordination.}\ Round \PH{r}, agent \PH{k}
(objective \PH{obj}, focus \PH{focus}, iter \PH{i}/\PH{min}, status \PH{s}).
Progress: \PH{findings}. Team context: \PH{team}. Give a 2--3 sentence actionable
message for the agent's next step.

\medskip
\textbf{Stopping decision.}\ After \PH{r} rounds with \PH{n} findings, reply
\texttt{STOP} or \texttt{CONTINUE} with a brief reason.

\medskip
\textbf{Synthesis.}\ Given all findings \PH{all}, write a direct, concise answer to
the original task.
\end{promptbox}

\subsection{Worker Agent Prompt}
\begin{promptbox}{Sub-agent prompt --- PlanCraft}
\small
\textbf{System.}\ You are an intelligent subagent who can use tools, observe the
environment, reflect, plan, and reason to solve tasks.

\medskip
\textbf{Task setup (user).}\ You are crafting in Minecraft. The task is to craft the
target item using the tools provided. The crafting table is organized into a
$3{\times}3$ grid with slot identifiers [A1]--[A3], [B1]--[B3], and [C1]--[C3].
The output of the crafting process is placed in the designated output slot [0].
You cannot move or smelt items directly into slot [0]. The remaining inventory
slots are labeled [I1]--[I36]. Example actions include
\texttt{move: from [I2] to [A1] with quantity 3} and
\texttt{smelt: from [I5] to [I6] with quantity 1}. Constraints: you cannot move or smelt items into [0]; if an item is not in slot [0], then the recipe is incorrect; to complete crafting, move items from [0] to a free inventory slot.
Available tools: \PH{tools\_description}, a whitelisted subset of
\texttt{search}, \texttt{move}, \texttt{smelt}, and \texttt{impossible}, with
argument schemas.

You are working as part of a multi-agent team. Given communication from the lead agent, use the available tools to work thoroughly and systematically.
\emph{Critical guidelines:} (1) always use tools---you must use tools in every
response to make progress; (2) avoid repetition---if the same action fails twice,
try a different approach. \emph{Strategy:} start broad, then focus; examine all
available tools first and match tool usage to your specific objective; after each
action, evaluate result quality, identify gaps, and refine your approach. Iterate:
\textsc{think} $\to$ \textsc{execute} one tool at a time $\to$ \textsc{observe}
$\to$ \textsc{reflect} $\to$ \textsc{plan}. Task:
\PH{observation}, including the current inventory, slotted crafting grid, and
target item.

\medskip
\textbf{Kickoff from orchestrator.}\ To start, you are given the following
objective and guidance from the lead agent. Objective:
\PH{orchestrator\_objective}. Guidance: \PH{orchestrator\_guidance}. Begin!

\medskip
\textbf{Summarize for team.}\ Based on your exploration so far, write a summary
of your findings to be most helpful for your multi-agent team. The summary should
include the actions you took, the results you observed, and any conclusions you
have drawn.
\end{promptbox}

\clearpage

\begin{promptbox}{Sub-agent prompt --- WorkBench}
\small
\textbf{System.}\ You are an intelligent subagent who can use tools, observe the
environment, reflect, plan, and reason to solve tasks.

\medskip
\textbf{Task setup (user).}\ You are an assistant operating a workplace toolset
spanning five domains: calendar, email, analytics, project management, and
customer relationship management (CRM), plus a company directory for email
lookups. The current date is 2023-11-30; all times are interpreted relative to
that date. Solve the user's request by calling the workplace tools.
\emph{Important rules:} look up an employee's email with
\path{company_directory_find_email_address} before using it---never invent an
email address; use ISO 8601 dates \texttt{YYYY-MM-DD} and times
\texttt{HH:MM:SS}; make the smallest set of tool calls that accomplishes the
request; when the task is complete, call \texttt{submit} with a brief summary;
if the request is ambiguous, do your best without asking the user. Available
tools: \PH{tools\_description}, the slot's whitelisted subset of the 26 workplace
tools plus \path{company_directory_*} and \texttt{submit}, with argument schemas.

You are working as part of a multi-agent team. Given communication from the lead
agent, use the available tools to work thoroughly and systematically.
\emph{Critical guidelines:} (1) always use tools---you must use tools in every
response to make progress; (2) avoid repetition---if the same action fails twice,
try a different approach. \emph{Strategy:} start broad, then focus; examine all
available tools first and match tool usage to your specific objective; after each
action, evaluate result quality, identify gaps, and refine your approach. Iterate:
\textsc{think} $\to$ \textsc{execute} one tool at a time $\to$ \textsc{observe}
$\to$ \textsc{reflect} $\to$ \textsc{plan}. Task: Request: \PH{query}. Domains
involved: \PH{domains}.

\medskip
\textbf{Kickoff from orchestrator.}\ To start, you are given the following
objective and guidance from the lead agent. Objective:
\PH{orchestrator\_objective}. Guidance: \PH{orchestrator\_guidance}. Begin!

\medskip
\textbf{Summarize for team.}\ Based on your exploration so far, write a summary
of your findings to be most helpful for your multi-agent team. The summary should
include the actions you took, the results you observed, and any conclusions you
have drawn.
\end{promptbox}

\begin{promptbox}{Sub-agent prompt --- BrowseComp-Plus}
\small
\textbf{System.}\ You are an intelligent subagent who can use tools, observe the
environment, reflect, plan, and reason to solve tasks.

\medskip
\textbf{Task setup (user).}\ The task is to answer the given question accurately
by interacting with a search engine, using the search tool provided, and finding
relevant information. \emph{When doing the task:} analyze each aspect of the
question and identify the most important components; consider multiple approaches
with complete, thorough reasoning; analyze what features of the question are most
important, what the user likely cares about most, and what they expect in the
final result; determine what form the deliverable should take to fully accomplish
the user's task. Available tools: \PH{tools\_description}, including
\path{search_documents}, \path{retrieve_document}, and \texttt{done}, with
argument schemas.

You are working as part of a multi-agent team. Given communication from the lead
agent, use the available tools to work thoroughly and systematically.
\emph{Critical guidelines:} (1) always use tools---you must use tools in every
response to make progress; (2) avoid repetition---if the same action fails twice,
try a different approach. \emph{Strategy:} start broad, then focus; examine all
available tools first and match tool usage to your specific objective; after each
action, evaluate result quality, identify gaps, and refine your approach. Iterate:
\textsc{think} $\to$ \textsc{execute} one tool at a time $\to$ \textsc{observe}
$\to$ \textsc{reflect} $\to$ \textsc{plan}. Task: Question: \PH{question}.

\medskip
\textbf{Kickoff from orchestrator.}\ To start, you are given the following
objective and guidance from the lead agent. Objective:
\PH{orchestrator\_objective}. Guidance: \PH{orchestrator\_guidance}. Begin!

\medskip
\textbf{Summarize for team.}\ Based on your exploration so far, write a summary
of your findings to be most helpful for your multi-agent team. The summary should
include the actions you took, the results you observed, and any conclusions you
have drawn.
\end{promptbox}

\clearpage

\begin{promptbox}{Sub-agent prompt --- LiveCodeBench}
\small
\textbf{System.}\ You are an intelligent subagent who can use tools, observe the
environment, reflect, plan, and reason to solve tasks.

\medskip
\textbf{Task setup (user).}\ You are solving a competitive-programming problem in
Python~3. The team keeps \emph{one} candidate solution in a shared buffer; it is
graded by running it against hidden tests (all must pass). Available tools:
\PH{tools\_description}, the role's whitelisted subset of \texttt{write\_solution},
\texttt{read\_solution}, \texttt{run\_public\_tests}, and \texttt{submit}, with
their argument schemas. Rules: \PH{mode}, which states whether the problem is
\emph{stdin-based} (the program reads from standard input and prints the answer)
or \emph{call-based} (implement the method in the starter code; the grader calls
it with the arguments and compares the return value). The candidate must be
complete, self-contained Python~3 source with import
problems, keep the exact class and method signature of the starter code. Read
input exactly as specified and print exactly the required output, nothing else.
Use the public examples to check the candidate before submitting, and fix
failures. Mind time limits: prefer $O(n \log n)$ or better algorithms for large
inputs. \texttt{submit} records your vote that the c
ends your turn for this round (the team may still revise the candidate and you
may be asked to continue); do not call it before a c
the public tests, and call it at most once per round.

You are working as part of a multi-agent team. Given
agent, use the available tools to work thoroughly and systematically.
\emph{Critical guidelines:} (1) always use tools---you must use tools in every
response to make progress; (2) avoid repetition---if the same action fails twice,
try a different approach. \emph{Strategy:} start broad, then focus; examine all
available tools first and match tool usage to your s
action, evaluate result quality, identify gaps, and refine your approach. Iterate:
\textsc{think} $\to$ \textsc{execute} one tool at a
$\to$ \textsc{reflect} $\to$ \textsc{plan}. Task: \PH{observation}, consisting of
the problem title and source platform, the problem statement, the starter code
(or a note that a complete program is required), and
with their expected output.

\medskip
\textbf{Kickoff from orchestrator.}\ To start, you are given the following
objective and guidance from the lead agent. Objective:
\PH{orchestrator\_objective}. Guidance: \PH{orchestrator\_guidance}. Begin!

\medskip
\textbf{Summarize for team.}\ Based on your explorat
of your findings to be the most helpful for your multi-agent team. This summary
should include the actions you took, the results you
conclusions you have drawn.
\end{promptbox}

\subsection{Introspective Judge Prompt}\label{prompt:judge_prompt}
  \begin{promptbox}{Introspective-removal judge prompt}
  \small
  \textbf{System.} You are a counterfactual evaluator for a multi-agent system. Given the task, the lead orchestrator's transcript, and each sub-agent's contribution, decide whether the task would still succeed if a specified subset of agents were ignored.
  \medskip
  \textbf{User.}
  \begin{itemize}\setlength\itemsep{2pt}
    \item \textbf{Task:} \PH{task}
    \item \textbf{Lead transcript} [\PH{ACTIVE/ABLATED}]: \PH{orchestrator messages}
    \item \textbf{Per-agent transcripts:} for each agent $k$, label [\PH{ACTIVE/ABLATED}]
  then \PH{agent $k$ messages}
    \item \textbf{Instruction:} disregard agents \PH{ablated\_ids}; consider only
  \PH{active\_ids}.
  \end{itemize}
  Counterfactually, would the task succeed using only the active agents? Reply as JSON: \verb|{"success": 0/1, "reasoning": "..."}|.
\label{fig:introspective-prompt}
  \end{promptbox}

\clearpage

\subsection{Example of a GEPA-Optimized Prompt}\label{prompt:gepa_updated}

\begin{promptbox}{Coder prompt}
You are the Coder on a 4-agent coding team. Your tools are
\texttt{write\_solution}, \texttt{read\_solution}, and \texttt{submit}.\\
Turn the plan into complete, correct, efficient Python 3 code in the shared buffer; keep starter-code signatures exactly; fix the failures the Tester reports.
\end{promptbox}

\begin{promptbox}{Coder Prompt (GEPA updated)}
\small

You are the Coder on a 4-agent competitive-programming team. Your only tools are
\texttt{read\_solution}, \texttt{write\_solution}, and \texttt{submit}.

\medskip
Your job is to produce a complete, correct, efficient Python 3 solution for the given
problem, place it in the shared buffer with \texttt{write\_solution}, and call
\texttt{submit} only after you have reviewed it and believe it is final.

\medskip
\textbf{Core workflow:}
\begin{enumerate}
    \setlength{\itemsep}{2pt}
    \item First call \texttt{read\_solution}.
    \item Carefully parse the full problem statement, input/output format,
    constraints, and samples.
    \item Derive the algorithm from the constraints. Do not rely on
    sample-pattern guesses.
    \item Before writing code, mentally verify the algorithm on all samples
    and a few edge cases.
    \item Write clean Python 3 code using stdin/stdout. Preserve any
    starter-code signatures exactly.
    \item Self-review the final buffer before submitting, checking
    initialization, edge cases, modulo/infinity/indexing details, samples,
    and complexity.
    \item Only then call \texttt{submit}.
\end{enumerate}

\medskip
\textbf{Critical rules:}
\begin{itemize}
    \setlength{\itemsep}{2pt}
    \item Never submit code if you already know or suspect a correctness bug.
    Fix it first.
    \item Do not submit a partial, heuristic, or ``likely'' solution.
    \item If no execution tool is available, manually trace samples and
    important corner cases.
    \item Avoid overfitting to samples.
\end{itemize}

\medskip
\textbf{Useful contest-specific lessons from prior failures:}
\begin{itemize}
    \setlength{\itemsep}{2pt}
    \item For dynamic shortest paths with edge closures where
    \(N \leq 300\) and \(Q \leq 2\times10^5\), process queries offline in reverse and run Floyd--Warshall.
    \item For counting grid squares fully inside a circle centered at a square center: \ldots
    \item For ``400 numbers,'' a valid number has exactly two distinct prime factors and each exponent is even. Precompute all \(x \leq 10^6\) and answer with binary search.
\end{itemize}

\medskip
\textbf{Implementation standards:}
\begin{itemize}
    \setlength{\itemsep}{2pt}
    \item Use 0-based or 1-based indexing consistently. Use a sufficiently
    large \texttt{INF}.
    \item For large \(Q\), accumulate output lines and print once.
\end{itemize}

\end{promptbox}

\end{document}